\documentclass[]{aastex631}

\begin{document}

\title{A Multiscale Analysis of an Interplanetary Shock from IMAP, SOLAR-1, and the L1 Constellation to Earth}

\correspondingauthor{Savvas Raptis}
\email{savvas.raptis@jhuapl.edu}
\email{savvasraptis@pm.me}
\author[0000-0002-4381-3197]{Savvas Raptis}
\affiliation{Johns Hopkins University Applied Physics Laboratory, Laurel, MD, USA}

\author[0000-0002-2425-7818]{Drew L. Turner}
\affiliation{Johns Hopkins University Applied Physics Laboratory, Laurel, MD, USA}

\author[0000-0002-0608-8897]{Domenico Trotta}
\affiliation{European Space Agency (ESA), European Space Astronomy Centre (ESAC), Camino Bajo del Castillo s/n, 28692, Villanueva de la Ca\~nada, Madrid, Spain}

\author[0000-0001-7024-1561]{Vassilis Angelopoulos}
\affiliation{University of California, Los Angeles, CA, USA}

\author[0000-0003-0452-8403]{James L. Burch}
\affiliation{Southwest Research Institute, San Antonio, TX, USA}

\author[0000-0003-2693-5325]{Dibyendu Chakrabarty}
\affiliation{Physical Research Laboratory, Ahmedabad, Gujarat, India}

\author[0000-0003-2134-3937]{Eric Christian}
\affiliation{NASA Goddard Space Flight Center, Greenbelt, MD, USA}

\author[0000-0002-0978-8127]{Christina M. S. Cohen}
\affiliation{California Institute of Technology, Pasadena, CA, USA}

\author[0000-0002-7341-2992]{Manuel E. Cuesta}
\affiliation{Department of Astrophysical Sciences, Princeton University, Princeton, NJ, USA}

\author[0000-0001-9323-1200]{Maher A. Dayeh}
\affiliation{Southwest Research Institute, San Antonio, TX, USA}

\author[0000-0002-2504-4320]{Robert W. Ebert}
\affiliation{Southwest Research Institute, San Antonio, TX, USA}
\affiliation{University of Texas at San Antonio, San Antonio, TX, USA}

\author[0000-0001-7952-8032]{Hameedullah A. Farooki}
\affiliation{Department of Astrophysical Sciences, Princeton University, Princeton, NJ, USA}

\author[0000-0002-0850-4233]{Joe Giacalone}
\affiliation{University of Arizona, Tucson, AZ, USA}

\author[0000-0001-9979-2164]{Matina Gkioulidou}
\affiliation{Johns Hopkins University Applied Physics Laboratory, Laurel, MD, USA}

\author[0000-0002-7572-4690]{Timothy Horbury}
\affiliation{Imperial College London, London, UK}

\author[0000-0002-0606-7172]{Immanuel C. Jebaraj}
\affiliation{Department of Physics and Astronomy, University of Turku, Turku, Finland}

\author[0000-0001-6160-1158]{David J. McComas}
\affiliation{Department of Astrophysical Sciences, Princeton University, Princeton, NJ, USA}

\author[0000-0002-2267-4140]{Connor O'Brien}
\affiliation{Johns Hopkins University Applied Physics Laboratory, Laurel, MD, USA}

\author{Srikar Paavan Tadepalli}
\affiliation{U R Rao Satellite Centre (URSC), Indian Space Research Organisation, Bengaluru, India}

\author[0000-0002-8111-1444]{Jamie S. Rankin}
\affiliation{Department of Astrophysical Sciences, Princeton University, Princeton, NJ, USA}

\author[0000-0003-1406-4200]{K. Sankarasubramanian}
\affiliation{U R Rao Satellite Centre (URSC), Indian Space Research Organisation, Bengaluru, India}

\author[0000-0002-3737-9283]{Nathan Schwadron}
\affiliation{Los Alamos National Laboratory, Los Alamos, NM 87545, USA}
\affiliation{Department of Astrophysical Sciences, Princeton University, Princeton, NJ, USA}
\affiliation{Space Science Center, University of New Hampshire, Durham, NH 03824, USA}

\author[0000-0002-5247-4107]{Bishwas L. Shrestha}
\affiliation{Department of Astrophysical Sciences, Princeton University, Princeton, NJ, USA}

\author[0000-0003-3255-9071]{Adam Szabo}
\affiliation{NASA Goddard Space Flight Center, Greenbelt, MD, USA}

\author[0000-0001-7188-8690]{Roy Torbert}
\affiliation{University of New Hampshire, Durham, NH, USA}

\author[0009-0007-6554-3611]{Dimitrios Vassiliadis}
\affiliation{NOAA/NESDIS/SWO, USA}

\author[0000-0002-7515-3285]{Sarah Vines}
\affiliation{Southwest Research Institute, San Antonio, TX, USA}

\author[0000-0002-8574-0416]{Hanying Wei}
\affiliation{University of California, Los Angeles, CA, USA}

\author[0000-0002-4313-1970]{Lynn B. Wilson III}
\affiliation{NASA Goddard Space Flight Center, Greenbelt, MD, USA}

\author[0000-0002-4642-6192]{Gary P. Zank}
\affiliation{The University of Alabama in Huntsville, Huntsville, AL, USA}

\author[0000-0002-4299-0490]{Lingling Zhao}
\affiliation{The University of Alabama in Huntsville, Huntsville, AL, USA}

\begin{abstract}
Collisionless shocks are primary sites of particle acceleration throughout the Universe. Transient, mesoscale structures forming around these shocks influence their energy conversion efficiency, yet remain poorly observed at interplanetary (IP) shocks: spacecraft sample the shock transition briefly, and single-point measurements cannot disentangle temporal evolution from spatial structure. Here, we examine a fast, CME-driven IP shock of 2026 January 19, observed by IMAP, SOLAR-1, Wind, ACE, DSCOVR, and Aditya-L1 at L1 and by ARTEMIS and MMS near Earth, to determine its spatial and temporal structure. At MHD scales, a pre-existing solar wind flux tube is overtaken by the shock as it crosses the L1 fleet. This inflates the apparent magnetic compression ratio far beyond the ideal strong-shock limit of 4, showing that upstream and downstream intervals can sample different plasma and bias single-spacecraft estimates. At kinetic scales, IMAP and SOLAR-1, separated by only $\sim7\,R_E$ ($\sim340$ ion inertial lengths), sample the same foreshock yet record localized structures that differ measurably; these are the first candidate signatures of large-scale foreshock transients at an IP shock. Further upstream, shocklets appear hundreds of $R_E$ ahead of the shock, consistent with steepening of ambient solar wind waves. Near Earth, the shock-to-flux-tube arrival lag increases to approximately six minutes, corresponding to a nominal sheath extension of $\sim50\,R_E$, indicating sheath expansion over $\sim200\,R_E$ of propagation. Together, these results show that the IP shock, its foreshock, and its sheath evolve across multiple scales from L1 to Earth, underscoring the need for multipoint measurements to resolve the underlying processes.\end{abstract}

\keywords{Shocks (2086) ---  The Sun (1693) --- Solar coronal mass ejections (310) --- Interplanetary shocks (829) --- Planetary bow shocks (1246) --- Heliophysics (2373)
 --- Space plasmas (1544) --- Plasma astrophysics (1261) --- Plasma physics (2089) --- Heliosphere (711)}

\section{Introduction}\label{sec:intro}

Collisionless shocks are ubiquitous in space and astrophysical plasmas, from planetary bow shocks and interplanetary (IP) shocks to supernova remnants \citep{treumann2009fundamentals,Balogh2013, Burgess2015}. In the absence of particle collisions, they convert directed flow energy into heating, electromagnetic fields, and a population of accelerated particles through wave-particle interactions, and are widely regarded as primary sites of particle acceleration in the Universe \citep{Blandford1978, Drury1983,kennel1985quarter,Giacalone2012,zank2015diffusive}. The efficiency of these processes depends strongly on the shock geometry, set by the angle $\theta_{Bn}$ between the shock normal and the upstream magnetic field. At quasi-parallel geometries ($\theta_{Bn} \lesssim 45^\circ$), reflected and backstreaming particles travel far upstream and drive an extended, variable foreshock \citep{Eastwood2005,burgess2012ion,wilson2016low,trotta2026energetic,jebaraj2026wave}. Within the foreshock region, transient structures form that scatter and energize particles efficiently, making quasi-parallel shocks especially effective accelerators over an extended spatial range \citep{Park2015,  Wilson2016a, liu2019relativistic,Turner2018, Giuffrida2022, Caprioli2023, jebaraj2024direct,Shi2025, Raptis2025a, Raptis2026b}. At quasi-perpendicular geometries, acceleration can be quite significant, though the mechanisms differ and the foreshock plays a less prominent role \citep{jokipii1987rate,Giacalone2005,Amano2020,Lindberg2024}.

Within the foreshock, transient mesoscale structures (tens to hundreds of ion inertial lengths) are increasingly recognized as central to shock energy conversion, particle acceleration, and space-weather coupling \citep{zhang_dayside_2022,Kajdi2024}. One family of these structures is intrinsically generated by the shock-foreshock system itself and requires no particular driver in the upstream flow: shocklets and short large-amplitude magnetic structures (SLAMS) are nonlinear structures that grow from steepening ultra-low-frequency waves \citep{Hoppe1981, Schwartz1992, Wilson2013a, collinson2023shocklets}, and high-speed jets that form downstream of the non-stationary quasi-parallel shock \citep{Hietala2009,Plaschke2018,Raptis2022a,Osmane2024,Krmer2024}. A second family instead requires an external trigger — typically a solar wind discontinuity convecting into the shock — including hot flow anomalies (HFAs) and foreshock bubbles (FBs) which are significantly larger in scale size \citep{thomas1988evolution, schwartz1985active, Omidi2010, turner2013first,Turc2025}. 

Such transients are known to be particularly efficient at scattering and accelerating particles which may exceed the Bohm limit \citep{Malkov2006}. At planetary shocks, they are known to perturb planetary magnetospheres while, for most, their scale has been found to grow with that of the host shock system \citep{Uritsky2014,Valek2017,Zhou2024,Raptis2026b}.

While these processes are routinely observed at planetary bow shocks, their counterparts at IP shocks remain scarce despite decades of near-Earth solar wind monitoring \citep{wilson2009low,trotta2023multi,hietala2024candidates}. In recent work we argued that this scarcity is largely observational rather than physical, although the statistically lower Mach numbers of IP shocks compared with planetary bow shocks may also limit foreshock development \citep{PrezAlanis2023,Trotta2025}: an IP shock sweeps past an in-situ spacecraft in seconds to minutes, so the nonlinear, spatially localized processes in the foreshock are sampled only briefly and rarely by more than one spacecraft at once. The relevant structures are also small and short-lived relative to the traversal, developing over a narrow growth zone \citep{Raptis2026a}. Consequently, capturing them may be a probabilistic problem, favoring fortuitous conjunctions of strategically-placed, well-instrumented spacecraft in the solar wind ahead of a strong shock.

Beyond their role as particle accelerators, IP shocks are of central
importance to space weather. The fastest and most energetic are driven by
coronal mass ejections (CMEs) and are the principal source of intense
geomagnetic storms \citep{Zhang2007,
Kilpua2015a,kilpua2021statistical,oliveira2023geoeffectiveness}. The
compressed, turbulent sheath that trails the shock, ahead of the driver
ejecta, is itself strongly geoeffective: its enhanced and fluctuating
magnetic field and dynamic pressure make sheaths a leading contributor to
storm main-phase development and to the storm-time response of the inner
magnetosphere \citep{Yermolaev2012, Lugaz2016, Kilpua2017}. Because space weather forecasting relies on upstream monitors at the first Lagrange point (L1) as proxies for the solar wind that subsequently reaches Earth, any evolution of the shock or its sheath over the intervening $\sim200\,R_E$ directly affects the accuracy of those predictions. How the shock and its sheath evolve between L1 and the Earth's magnetosphere can therefore have a direct impact on space weather.

A common limitation affects both the study of transient-process formation and the assessment of space-weather impact. Collisionless shocks are open systems, continuously driven by an upstream medium that carries both small-amplitude fluctuations and finite-amplitude structures such as discontinuities and flux tubes into the shock transition. Yet shock parameters such as the compression ratio, Mach number, and $\theta_{Bn}$ are routinely derived from a single spacecraft, under the assumption that the sampled upstream and downstream intervals represent the same uniform plasma. When the inflow is structured on scales comparable to or larger than the analysis window, this assumption breaks down, and a single monitor cannot separate the shock from the particular medium it happens to be crossing. Resolving this ambiguity, and following how the shock and its surroundings evolve across scales, requires distributed, multipoint measurements in the near-Earth solar wind that have until recently been unavailable \citep{Lugaz2024,palmerio2024mesoscale}.

The arrival of NASA's Interstellar Mapping and Acceleration Probe (IMAP), and NOAA's Space weather Observations at L1 to Advance Readiness - 1 (SOLAR-1) near L1, alongside the established solar wind monitors, now provides such coverage. In this work, we use this L1 constellation, together with near-Earth observations from Acceleration, Reconnection, Turbulence and Electrodynamics of Moon's Interaction with the Sun (ARTEMIS) and Magnetospheric Multiscale (MMS), to follow a fast, CME-driven, IP shock across multiple scales during its passage on 2026 January 19. We show that (i) at large, MHD scales, a flux tube overtaken by the shock biases single-spacecraft compression ratios well beyond the strong-shock limit, so that the immediately inferred shock characteristics at one L1 monitor do not describe the shock elsewhere or at Earth; (ii) at kinetic scales, the closely spaced IMAP and SOLAR-1 spacecraft resolve a variable foreshock that hosts both steepening shocklets arising from underlying solar wind fluctuations far upstream and candidate signatures of large-scale foreshock transients, the latter differing measurably between the two spacecraft despite their small separation; and (iii) at global scales (L1 to Earth), the same flux tube serves as a timing marker that tracks the growth of the CME sheath between L1 and MMS. Together, these results follow the IP shock, its foreshock, and its sheath as one system evolving across kinetic, MHD, and global scales, and show that multipoint measurements are necessary to resolve their underlying physics and advance our understanding of particle energization at shocks, a primary objective of the IMAP mission \citep{Cohen2026}.

\section{Data and Methods}\label{sec:data}

The event examined here was monitored by a constellation of six spacecraft at L1: NASA's Interstellar Mapping and Acceleration Probe (IMAP) \citep{McComas2018, McComas2025}, SOLAR-1 \citep{Vassiliadis2026}, Wind \citep{Ogilvie1997,Wilson2021}, the Advanced Composition Explorer (ACE) \citep{Stone1998}, the Deep Space Climate Observatory (DSCOVR) \citep{Burt2012}, and Aditya-L1 \citep{Seetha2017,Parate2025,Sankarasubramanian2025}. The shock's subsequent arrival in the near-Earth environment was observed by ARTEMIS \citep{Angelopoulos2008,Angelopoulos2011} in lunar orbit and by the MMS mission \citep{Burch2016} upstream of Earth's bow shock. Unless otherwise stated, all data are presented in the Geocentric Solar Ecliptic (GSE) coordinate system, in which $X$ points from Earth to the Sun, $Z$ is normal to the ecliptic plane and positive northward, and $Y$ completes the right-handed set toward dusk. Spatial scales are given in Earth radii ($R_E$) and, where kinetic scales are relevant, normalized to a reference ion inertial length, $d_i = c/\omega_{pi} \approx 228/\sqrt{n_i}\,\mathrm{km}$ for proton density $n_i$ in $\mathrm{cm^{-3}}$.  Where minimum variance analysis (MVA) is applied, we work in the associated $LMN$ coordinate system: $\hat{\mathbf{L}}$ and $\hat{\mathbf{M}}$ span the maximum- and intermediate-variance directions, respectively, while $\hat{\mathbf{N}}$ is the minimum-variance direction, taken as the approximate wave-propagation or boundary-normal direction \citep{Sonnerup1967,khrabrov1998error}.

Our analysis rests primarily on magnetic field measurements, available from all eight spacecraft, complemented by plasma moments from the two most closely spaced L1 monitors (and, in Appendix~\ref{sec:appendix_speed}, from Wind and Aditya-L1). At L1, the IMAP magnetometer \citep{Horbury2026} provides survey data at $0.5$~s resolution ($2$~Hz) and burst data at $64$~Hz; IMAP spins with a $15$~s period, so a spin-tone signal near ${\sim}0.067$~Hz falls within the spectral range analyzed below, though it should not significantly affect our power spectral density (PSD) ratios. The SOLAR-1 magnetometer \citep{Torbert2026} provides data at $64$~Hz; the Wind Magnetic Field Investigation \citep{Lepping1995} samples at $\sim 11$~Hz; the ACE \citep{Smith1998} and DSCOVR \citep{Loto2022} magnetometers at $1$~Hz; and the Aditya-L1 fluxgate magnetometer \citep{Yadav2025} at $0.1$~Hz. Near Earth, we use the MMS fluxgate magnetometer \citep{Russell2016} in survey mode (0.0625~s), with $128$~Hz burst data where available, and the ARTEMIS fluxgate magnetometer \citep{Auster2008} at $\sim 4$-s spin resolution. The $64$~Hz IMAP burst and SOLAR-1 data resolve ion-kinetic scales in the foreshock and underpin the shocklet and transient analysis presented below. Plasma moments are available at $1$-minute resolution from IMAP, using the Solar Wind and Pickup Ion instrument (SWAPI) \citep{Rankin2025}, and from SOLAR-1, using the Solar Wind Plasma Sensor (SWiPS) \citep{Ebert2026}; for IMAP we additionally use the $12$~s pseudo-moments from the IMAP Active Link for Real-Time (I-ALiRT) system \citep{Lee2025} to resolve finer-scale signatures. As IMAP and SOLAR-1 were early in operations during this event, their plasma products should be treated with caution. Finally, we characterize the associated geomagnetic storm using the SuperMAG ring current index (SMR) \citep{Newell2012}, which is equivalent to the Sym-H index \citep{iyemori1990storm}. SMR is an average index based on all available ground stations located between $-50^{\circ}$ and $+50^{\circ}$ magnetic latitude (MLat).

We identify the local shock ramp at each spacecraft using the gradient in $|B|$ and characterize the shock using complementary single- and
multi-spacecraft techniques. For the multi-spacecraft timing analysis \citep{paschmann1998analysis,ToyEdens2025}, we evaluate the GSE positions of the six L1 monitors at their respective ramp times. Assuming a planar front propagating at a constant normal speed, the crossing times satisfy $t_i=t_0+\mathbf{r}_i\cdot\mathbf{s}$, where $\mathbf{s}=\hat{\mathbf{n}}_{\rm timing}/V_{\rm sh}$ is the three-component shock-slowness vector. We determine $\mathbf{s}$ with an unweighted least-squares fit that minimizes
\begin{equation}
\chi^2(\mathbf{s})=\sum_{i=1}^{6}\left[(t_i-\bar t)-(\mathbf{r}_i-\bar{\mathbf{r}})\cdot\mathbf{s}\right]^2,
\label{eq:timing_objective}
\end{equation}
where the bars denote averages over the six spacecraft, which is equivalent to fitting $t_0$ as a free parameter so that no spacecraft serves as a reference. We obtain the shock normal as $\hat{\mathbf{n}}_{\rm timing}=\mathbf{s}/|\mathbf{s}|$ and the shock speed as $V_{\rm sh}=|\mathbf{s}|^{-1}$, respectively. The axial sign is chosen so that the reported normal points Earthward. The method and its caveats are summarized in Appendix~\ref{sec:appendix_diagnostics}.

Single-spacecraft normals follow from magnetic coplanarity \citep{colburn1966discontinuities} and, near the shock where plasma moments are available for IMAP and SOLAR-1, we make additional use of the mixed-mode coplanarity (MX3) \citep{abraham1972determination,Vias1986,Trotta2022b}. The shock-normal angle $\theta_{Bn}$ is the angle between a local normal to the shock front and the upstream field, and we report it against both the coplanarity-derived normal and the multi-spacecraft timing one. Magnetic compression ratios are computed from the medians of $|B|$ in upstream and downstream windows and from the downstream peak of $|B|$. To quantify the uncertainty we also calculate the normal using timing by leaving one spacecraft out, obtaining a distribution of normals from which we compute quantiles as an estimate of variability.

Magnetic field PSDs of the field magnitude $|B|$ are estimated with Welch's method \citep{Welch1967} over sliding windows and therefore capture compressive (field-magnitude) fluctuations; inter-spacecraft PSD ratios are formed on a common frequency grid. For comparison, we also compute the trace PSD, $P_{xx}+P_{yy}+P_{zz}$, with identical windows and processing (Appendix~\ref{sec:appendix_spectra}). Separately, to identify compressive and transient structures jointly in time and frequency, we compute the complex-Morlet wavelet trace power, which sums the power in all three field components and therefore includes both compressive and transverse fluctuations \citep{torrence1998practical}. Upstream oscillatory packets are examined with MVA on the bandpass filtered magnetic field data, from which we report the ellipticity. For visualization, time intervals are also expressed as a nominal convected-distance scale using a reference speed. This conversion is used only as a proxy for distance from the shock and does not assume that the observed fluctuations satisfy the Taylor hypothesis; in particular, it neglects intrinsic wave propagation and assumes a stationary shock. Distances are additionally normalized by the ion inertial length ($d_i$) for a reference density. The reference speed and density adopted for each distance axis are stated with the corresponding figure, while for Earth radius we use $R_E=6371\,\mathrm{km}$.

\section{The 2026 January 19 Event}\label{sec:overview}

\begin{figure*}[ht]
    \centering
{\includegraphics[width=1\textwidth]{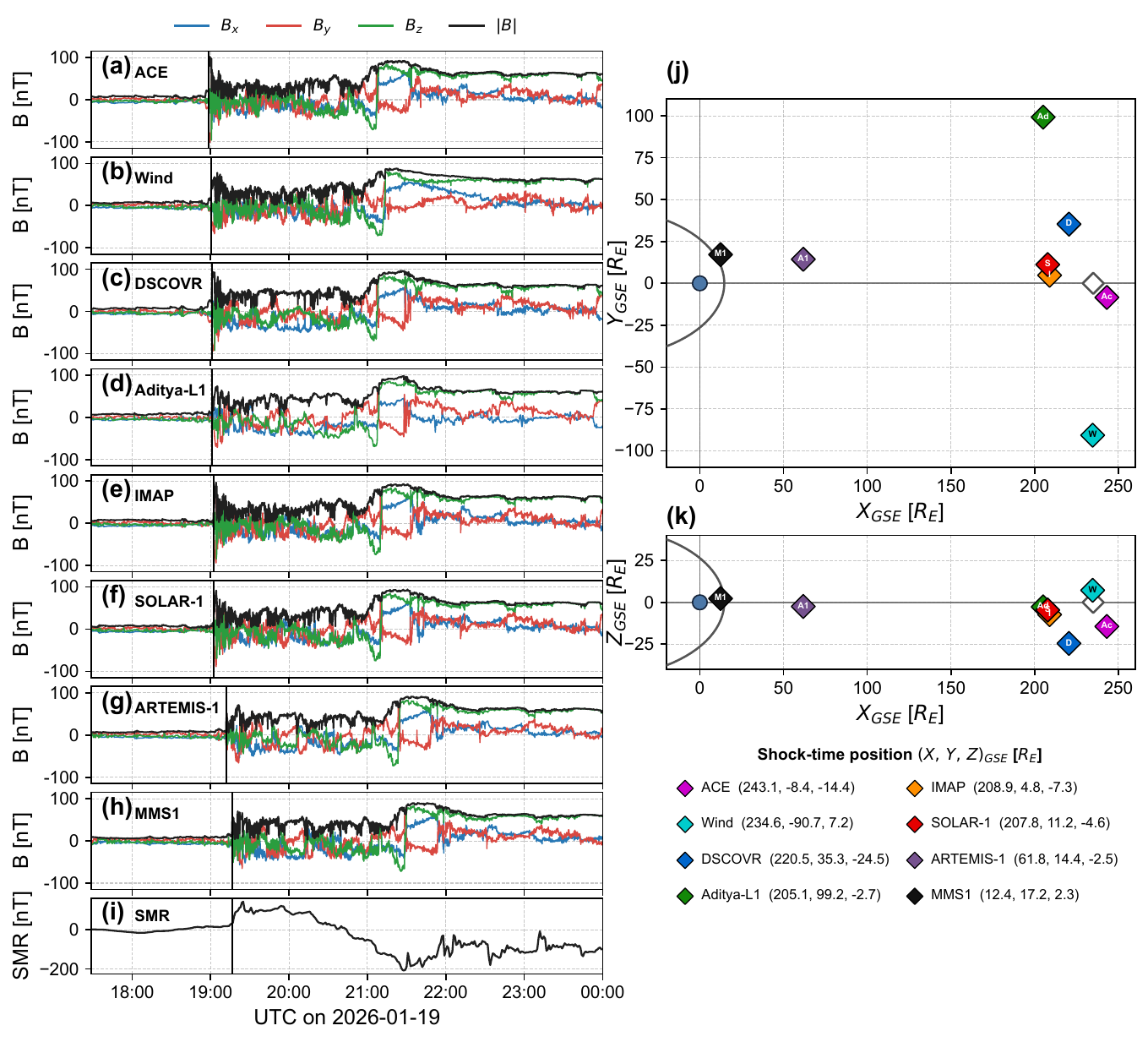}}\caption{{Multi-spacecraft overview of the 19 January 2026 interplanetary shock. Panels (a)--(h) show the GSE magnetic field components and magnitude observed by ACE, Wind, DSCOVR, Aditya-L1, IMAP, SOLAR-1, ARTEMIS-1, and MMS1, respectively, between 17:28:45 and 24:00:00 UTC. Vertical black lines mark the local shock ramps. Panel (i) shows the SMR index, with the vertical line marking the MMS1 shock arrival. Panels (j) and (k) show the spacecraft locations at their respective shock times in the GSE $X$--$Y$ and $X$--$Z$ planes. The blue circle marks Earth, the open diamond marks the nominal L1 point, and the gray curve shows an average bow-shock model \citep{farris1994determining}. 
The location-plot legend provides each spacecraft position in GSE coordinates, rounded to one decimal place.} \label{fig:1}}
\end{figure*}

Before shock-aligning the measurements, we establish the event, its geometry, and its geoeffective context. On 2026 January 19, a fast CME-driven IP shock swept across IMAP and the L1 constellation \citep{cuesta2027correlation} and reached the near-Earth spacecraft about 10--14 minutes later. Figure~\ref{fig:1} shows the magnetic field at all eight spacecraft, ordered by the local shock-ramp arrival time, together with the geomagnetic response and the spacecraft configuration.

The shock crossed the six L1 monitors in the order ACE, Wind, DSCOVR, Aditya-L1, IMAP, and SOLAR-1 in less than four minutes (Fig.~\ref{fig:1}a--f), and reached ARTEMIS and MMS near Earth about 10 and 14 minutes after IMAP, respectively (Fig.~\ref{fig:1}g,h). A planar timing fit to the six L1 crossings yields a global normal close to the Sun-Earth line ($\hat{\mathbf n}_{\rm timing}=(-0.964,-0.127,-0.235)$) and a shock speed of $\approx 850\,\mathrm{km/s}$ (820--950~km/s when leaving out one spacecraft at a time; Appendix~\ref{sec:appendix_diagnostics}), indicating a fast shock propagating nearly radially.

The downstream proton speed at Wind and Aditya-L1 ($\approx900$--$1000\,\mathrm{km/s}$; Figure~\ref{fig:speed}), however, approaches or exceeds this timing speed. Timing measures the propagation speed directly, but over the $\sim40\,R_E$ extent of the L1 fleet along the flow it is sensitive to the assumed normal, to inclination or rippling of the front, and to ramp-identification errors (Appendix~\ref{sec:appendix_diagnostics}). Over the longer baseline to ARTEMIS and MMS, the arrival times imply mean speeds of $\approx1300$--$1600\,\mathrm{km/s}$, and mass-flux conservation at Wind gives $\approx1400\,\mathrm{km/s}$ when the flux-tube plasma directly ahead of the ramp is taken as the upstream state (Appendix~\ref{sec:appendix_speed}). These differences likely reflect both geometric variability of the front and its interaction with the flux tube (Section~\ref{sec:mhd}); we therefore use the six-spacecraft timing speed as a reference for converting time to distance.

The upstream moments at IMAP and SOLAR-1 place the shock well above both the Alfv\'enic and the fast-magnetosonic thresholds. Using the timing solution, we obtain fast-magnetosonic Mach numbers $M_f\approx2.3$--$4.0$ (IMAP) and $\approx2.7$--$3.1$ (SOLAR-1) and Alfv\'en Mach numbers of $M_A\approx4$--$6$ (IMAP) and $\approx9$--$11$ (SOLAR-1); the faster local speeds discussed above would raise these values significantly. Although the exact values vary, the shock is unambiguously super-Alfv\'enic and supercritical (i.e., above the Mach number threshold at which resistivity alone cannot provide the required dissipation and ion reflection takes place) at both spacecraft, placing the shock in the regime in which an extended quasi-parallel foreshock and efficient particle acceleration are expected. We note that although the Mach numbers from the two spacecraft differ, the broader upstream magnetic field and plasma properties measured by IMAP and SOLAR-1 are in generally good agreement (see Appendix~\ref{sec:appendix_broader} and Figure~\ref{fig:extended_upstream_context}).

Although a single shock passed the fleet, the standard per-spacecraft diagnostics \citep{neugebauer2005multispacecraft} are strikingly inconsistent (Figure~\ref{fig:diagnostics}; full numerical values in Appendix Table~\ref{tab:l1_mc_appendix}). At ACE, Wind, and DSCOVR the magnetic compression ratio is ordinary,
$B_d/B_u \approx 3$--$5$, and the coplanarity and timing obliquities agree to
within about ten degrees. Aditya-L1 shows an ordinary compression but an
elevated peak ratio and a coplanarity normal that departs sharply from the timing
one ($\theta_{Bn}^{\mathrm{MC}} \approx 23^{\circ}$ against
$\theta_{Bn}^{\mathrm{timing}} \approx 70^{\circ}$). At IMAP and SOLAR-1, by contrast, the apparent compression is far larger ($B_d/B_u\approx23$ and $\approx8$, with peak ratios even larger), and the two obliquity estimates diverge, for example $\theta_{Bn}^{\rm MC}\approx36^{\circ}$ against $\theta_{Bn}^{\rm timing}\approx63^{\circ}$ at IMAP. These apparent magnetic compression ratios greatly exceed the ideal strong-shock reference value of 4 (the density compression limit for a polytropic index $\gamma=5/3$, which increases for smaller $\gamma$) and, together with the multipoint observations, indicate that the selected intervals sample different plasma structures \citep{moreland2023variability}. Figure~\ref{fig:diagnostics} summarizes this pattern: the three monitors where the flux tube remains partly upstream (ACE, Wind, DSCOVR) show ordinary compression and similar obliquity estimates, whereas the three where it has entered the downstream interval (Aditya-L1, IMAP, SOLAR-1) show inflated peak or median ratios and divergent obliquity estimates. The following section focuses on identifying the cause of this effect.

\begin{figure*}[ht]
    \centering
    {\includegraphics[width=0.75\textwidth]{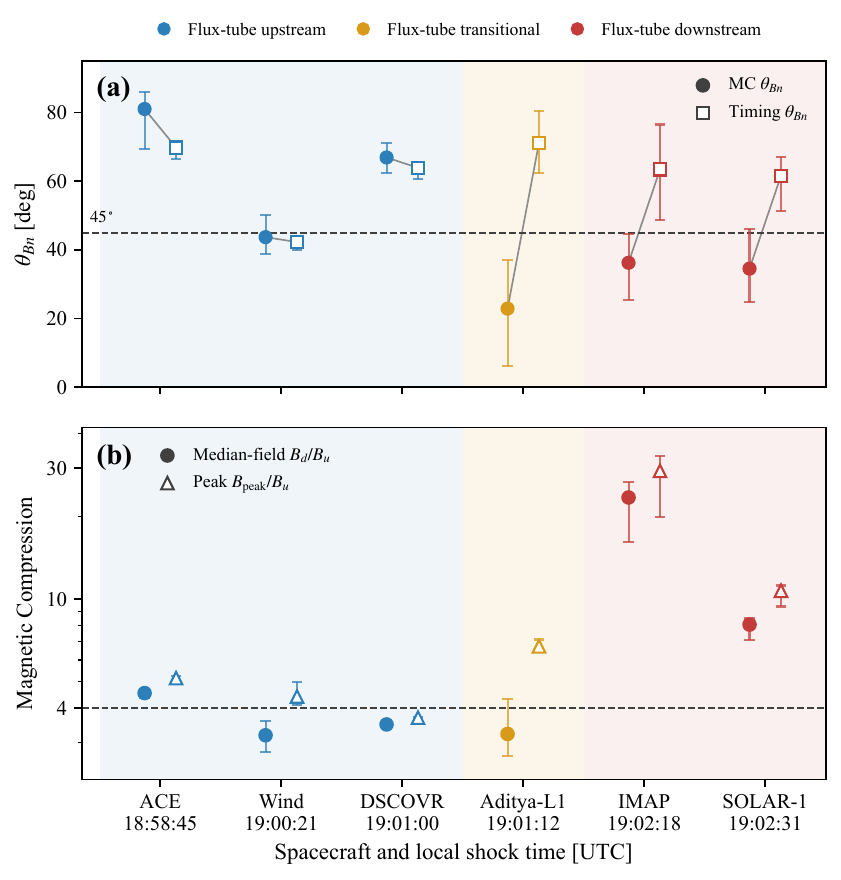}}
    \caption{Shock-diagnostic summary for the six L1 spacecraft, ordered by local shock arrival. (a)~Shock obliquity $\theta_{Bn}$ from magnetic coplanarity (MC) and from the common six-spacecraft timing normal ($\hat{\mathbf n}_{\rm timing}=(-0.964,-0.127,-0.235)$), with 16th--84th percentile sensitivity bars from varied upstream/downstream windows (see Figure~\ref{fig:2}). The horizontal line marks $\theta_{Bn}=45^{\circ}$. (b)~Magnetic compression ratio: median ($B_d/B_u$) and peak ($B_{\rm peak}/B_u$) values, with the ideal strong-shock reference of~4 (for $\gamma=5/3$) marked. Background shading separates the three spacecraft at which the solar wind flux-tube lies partly upstream of the ramp (ACE, Wind, DSCOVR) from those at which it lies downstream (IMAP, SOLAR-1), with Aditya-L1 as transitional. At ACE the median compression of $4.5$ slightly exceeds the limit, consistent with a modest contribution from the flux-tube edge entering the downstream window. Full numerical values, including coplanarity normal vectors, appear in Appendix Table~\ref{tab:l1_mc_appendix}.}
    \label{fig:diagnostics}
\end{figure*}

\section{MHD Scales: The Shock and Flux-Tube Bias}\label{sec:mhd}

We first examine the shock at scales much larger than the ion kinetic length (thousands of $d_i$, or tens of $R_E$), where an MHD description and single-shock jump conditions are normally applied.

\begin{figure*}[ht]
    \centering
{\includegraphics[width=1\textwidth]{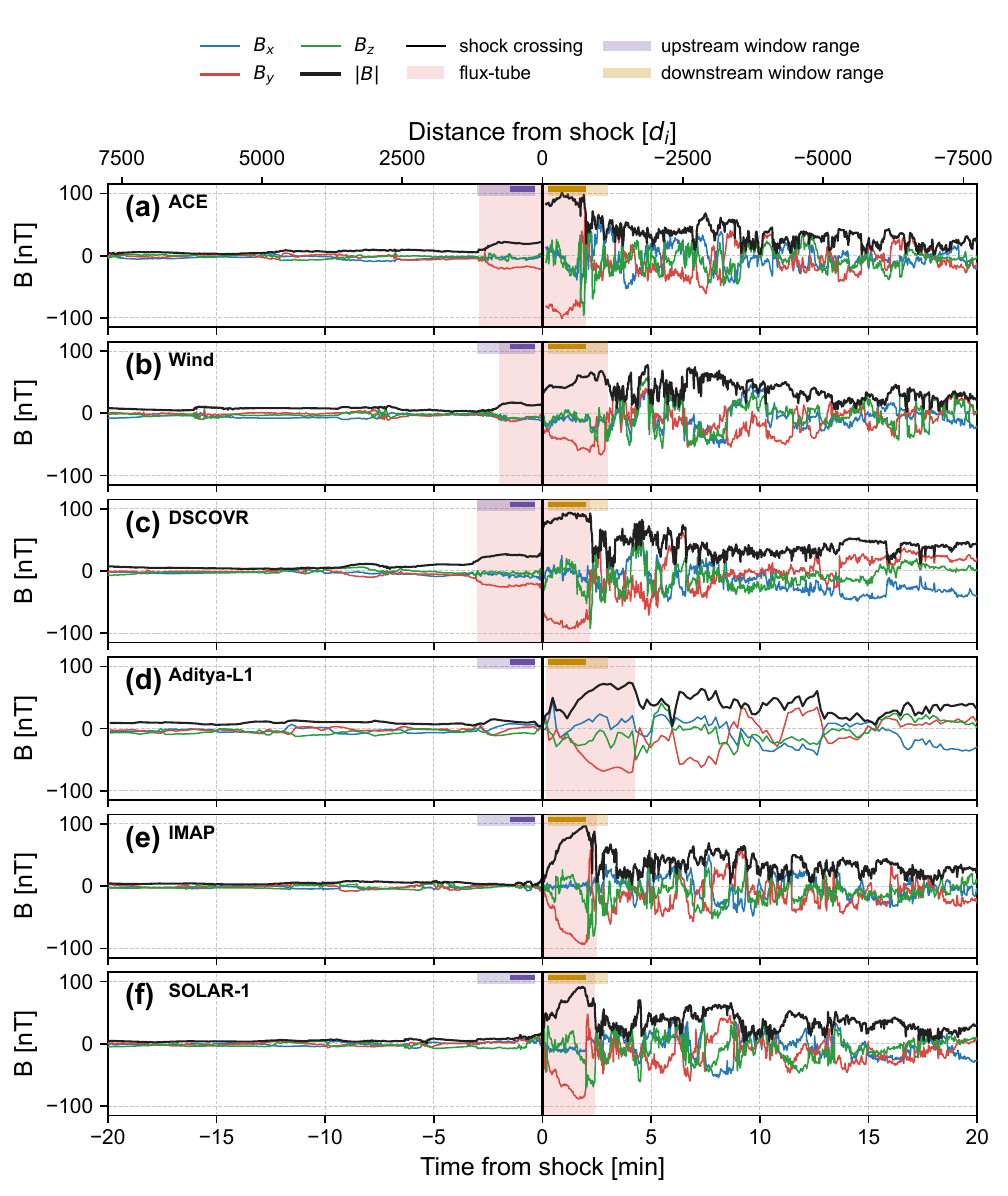}}\caption{Shock-aligned magnetic field data at L1. Panels (a)--(f) show the GSE magnetic field components and magnitude measured by ACE, Wind, DSCOVR, Aditya-L1, IMAP, and SOLAR-1, ordered by local shock arrival. Time is measured relative to each local shock ramp (vertical black lines). Red shading marks the mission-specific broad flux-tube envelope, which lies partly upstream at ACE, Wind, and DSCOVR but downstream at Aditya-L1, IMAP, and SOLAR-1. Purple and gold bars show the full upstream and downstream domains sampled by the window ensemble. The light bars show the full upstream ($-180$ to $-20$ s) and downstream ($+15$ to $+180$ s) window ranges; the darker bars mark the nominal averaging windows ($-90$ to $-20$ s and $+15$ to $+120$ s, respectively). The top axis gives an approximate distance from the shock in ion inertial lengths using $V_{\rm sh}=850\,\mathrm{km/s}$ and $n_i=3\,\mathrm{cm^{-3}}$. Because the downstream density differs from the upstream value used for the conversion, the downstream \(d_i\) axis is illustrative only. See also Figure~\ref{fig:diagnostics} for the corresponding diagnostic summary.\label{fig:2}}
\end{figure*}

The inconsistent diagnostics shown in Figure~\ref{fig:diagnostics} have a single, identifiable cause: a magnetic flux tube that the shock overtakes as it crosses the L1 constellation. Figure~\ref{fig:2} shock-aligns the L1 magnetic field and marks this structure (red envelope). The envelope is defined by the interval of sustained field-magnitude enhancement and coherent $B_y$ rotation that stands above the ambient upstream level at each spacecraft; its edges are placed where the magnitude returns to the local baseline. At ACE, Wind, and DSCOVR the structure lies just upstream of the ramp, whereas at Aditya-L1, IMAP, and SOLAR-1 it lies downstream, showing that the faster shock swept through it during the four minutes it took to cross the fleet. Because the shock overtakes the structure progressively across the constellation, the fraction that appears upstream versus downstream shifts from one spacecraft to the next, accounting for the different apparent durations of the envelope at each monitor. Because the structure does not need to be a straight cylinder, a meandering flux tube with a slab-like component, crossed at different distances from its axis by each spacecraft, can also contribute to these differences.

Along the Sun-Earth line the structure spans roughly $20$--$40\,R_E$ ($1000$--$2000\,d_i$) at each spacecraft, and because it is seen across the entire L1 fleet it must extend at least $\sim200\,R_E$ in $Y_{\rm GSE}$ and $\sim30\,R_E$ in $Z_{\rm GSE}$. It is therefore considered a mesoscale structure by solar wind terminology \citep{viall2021mesoscale}, larger than the terrestrial magnetosphere yet smaller than the IP shock itself. Its coherent field rotation and enhanced magnitude are consistent with a solar wind flux rope of the kind reported in the ambient solar wind \citep{borovsky2008flux, ruan2009multiple,blanco2019multispacecraft,shin2026statistical}.

The presence of this flux tube is what biases the single-spacecraft parameters. At IMAP and SOLAR-1 the interval immediately downstream of the ramp falls inside the high-field flux tube rather than in nominally shocked solar wind, so the measured $B_d/B_u$ is inflated to $\approx8$--$23$ (Figure~\ref{fig:diagnostics}b; Table~\ref{tab:l1_mc_appendix}). The same bias pushes the coplanarity normal away from the fleet consensus and opens the gap between $\theta_{Bn}^{\rm MC}$ and $\theta_{Bn}^{\rm timing}$ at these spacecraft. At Aditya-L1, where the shock has only just overtaken the flux tube, an ordinary median compression but an elevated peak ratio is observed, consistent with the flux tube entering the downstream averaging window. At the leading monitors, with the tube still partially upstream, both the compression and the geometry are normal. We stress that the inflated ratios are a direct consequence of the window placement: with careful manual selection of narrower intervals that exclude the flux-tube measurements, the compression at IMAP and SOLAR-1 can be reduced substantially. Ratios above 4 can also arise physically, e.g., from a variable polytropic index \citep{Jiao2026} or turbulence \citep{Nakanotani2026}, but here the large ratios originate from upstream and downstream intervals that sample different plasma. This, however, only reinforces the broader point. The ``correct'' upstream and downstream intervals are not self-evident from a single spacecraft's time series, and automated shock-characterization methods, which must choose windows algorithmically, are especially vulnerable to contamination by mesoscale structure of this kind \citep{oliveira2023geoeffectiveness,moreland2023variability,PrezAlanis2023,Trotta2025,cuesta2027correlation}.

The multi-spacecraft timing normal, insensitive to the plasma measured locally by a single spacecraft, provides an overall more reliable geometry here; the per-spacecraft coplanarity normals do not. However, this normal represents the global rather than the local shock geometry, and its orientation is only approximately constrained because the front is not strictly planar across the fleet (Section~\ref{sec:overview}). The timing baseline spans the $\sim200\,R_E$ extent of the L1 fleet in $Y_{\rm GSE}$, so the resulting normal averages the shock geometry over that mesoscale aperture. Constraining the geometry at larger scales, such as the full angular extent of the CME-driven shock, would require spacecraft separated by hundreds of $R_E$ or more beyond the current L1 constellation. More importantly, this event shows that a mesoscale structure can bias an MHD-scale shock inference, and that the parameters read off at one L1 monitor, including the compression, Mach number, and obliquity, may not describe the shock elsewhere in the fleet or by the time it reaches Earth.

\section{Foreshock Structure From Mesoscales to Kinetic Scales}\label{sec:kinetic}

\begin{figure*}[ht]
    \centering
{\includegraphics[width=1\textwidth]{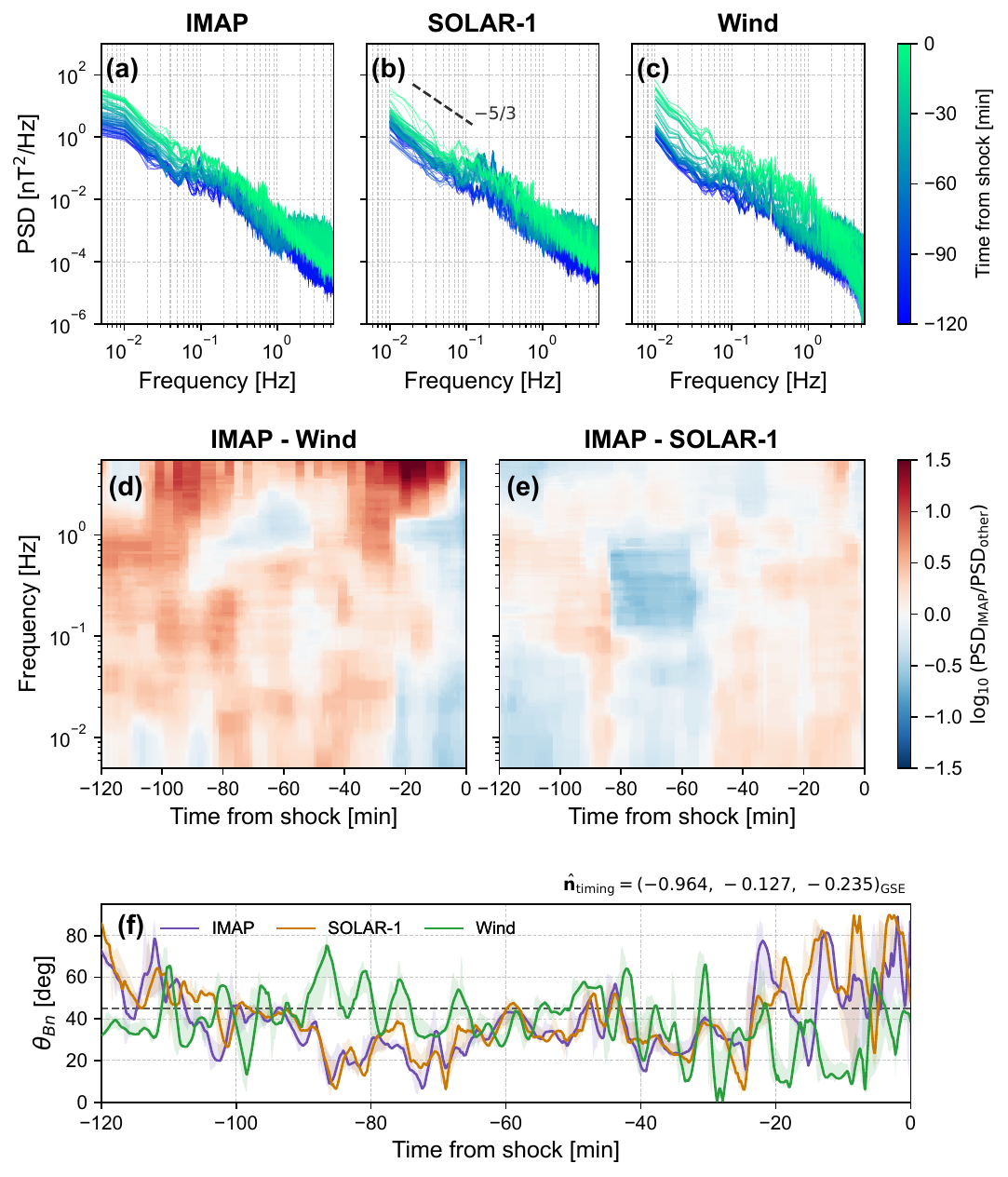}}\caption{Evolution of upstream magnetic field fluctuations and shock obliquity. Panels (a)--(c) show power spectral densities (PSD) of $|B|$ at IMAP, SOLAR-1, and Wind, calculated from 30-min windows advanced in 1-min steps and colored by the window endpoint relative to each local shock. The dashed segment in panel (b) is an $f^{-5/3}$ reference line. Panels (d) and (e) show the matched logarithmic power ratios $\log_{10}(P_{\rm IMAP}/P_{\rm Wind})$ and $\log_{10}(P_{\rm IMAP}/P_{\rm SOLAR-1})$, respectively; red therefore denotes stronger $|B|$ power at IMAP. These maps use a 21-bin median along log-frequency followed by a 5-min trailing median in time. Panel (f) shows $\theta_{Bn}$ calculated from each spacecraft's trailing 120-s magnetic field and the common six-spacecraft timing normal $\hat{\mathbf n}_{\rm timing}=(-0.964,-0.127,-0.235)_{\rm GSE}$. Shading spans the 16th--84th percentiles obtained from six leave-one-spacecraft-out timing normals and magnetic field averaged over 30--300 s. Therefore, strictly speaking, it represents methodological sensitivity rather than statistical confidence. The horizontal dashed line marks $\theta_{Bn}=45^{\circ}$.} \label{fig:3}
\end{figure*}

Having established that a distinct flux tube can distort the MHD-scale picture, we turn to the foreshock at ion-kinetic scales, where the wave field and shocklets are resolved, and to mesoscale transient structures near the shock. IMAP and SOLAR-1 are central here: closely spaced and carrying $64$~Hz magnetometers, they resolve the upstream foreshock in detail, and, with the flux tube already downstream of the shock, they sample the upstream foreshock without any external disturbance seen at the other monitors. We descend through this foreshock in three steps, from its broad wave-power evolution (Fig.~\ref{fig:3}), to discrete steepened wave packets, or shocklets, far upstream (Fig.~\ref{fig:4}; subsection \ref{sec:shocklets}), and finally to localized transient candidates near the shock (Fig.~\ref{fig:5}; subsection \ref{sec:transients}).

Figure~\ref{fig:3} tracks the evolution of upstream magnetic power at IMAP, SOLAR-1, and Wind over the two hours preceding each local shock crossing. Wind is included because it offers the highest magnetic-field cadence among the remaining L1 monitors and provides a comparison with the stronger upstream fluctuations observed at IMAP and SOLAR-1. At all three spacecraft, the $|B|$ power spectrum steepens and intensifies as the shock approaches (Fig.~\ref{fig:3}a--c), following an approximately $f^{-5/3}$ slope across the MHD inertial range, characteristic of solar wind turbulence \citep{Bruno2013, howes2015dynamical,park2025modification}. However, there are significant differences between each spacecraft. 

The logarithm of the IMAP-to-Wind power spectral density ratio (Fig.~\ref{fig:3}d) is positive across much of the interval and grows toward the shock at high frequencies, showing that IMAP develops a more compressive foreshock with stronger high-frequency activity. The IMAP-to-SOLAR-1 ratio (Fig.~\ref{fig:3}e) remains near zero, indicating comparable power and confirming that these two closely spaced spacecraft sample essentially the same foreshock, apart from local differences. We focus on $|B|$ because it isolates the compressive fluctuations of steepening foreshock waves, whereas the trace PSD is dominated by transverse, largely incompressible fluctuations of the ambient solar wind, whose ULF power is comparable at all three spacecraft; at these frequencies the IMAP fluctuations are thus roughly twice as compressive as those at Wind (Appendix~\ref{sec:appendix_spectra}).

The local geometry is correspondingly variable. The obliquity $\theta_{Bn}$, computed from the common timing normal and each spacecraft's trailing field (Fig.~\ref{fig:3}f), swings repeatedly across the $45^{\circ}$ quasi-parallel/quasi-perpendicular boundary at all three spacecraft throughout the interval. For a high-Mach shock this variability is expected: the upstream field, and with it the instantaneous connection to the shock, changes on timescales far shorter than the crossing, so a single $\theta_{Bn}$ value cannot represent the shock the particles actually experience. We therefore suggest that the sustained, shock-approaching foreshock wave field is a robust indicator of prolonged quasi-parallel connection, since it reflects the cumulative particle dynamics upstream rather than the instantaneous geometry at a single time. On this measure IMAP and SOLAR-1, near the shock, appear more persistently connected to quasi-parallel portions of the shock than Wind. We stress that this ordering is the opposite of what the ramp-time diagnostics in Table~\ref{tab:l1_mc_appendix} suggest, where Wind has the smallest $\theta_{Bn}^{\rm timing}$ of the fleet and yet the weakest compressive and high-frequency wave activity. This is the inconsistency the preceding argument anticipates, i.e., that an obliquity evaluated over a single interval at the ramp does not need to reflect the connection history that built the foreshock. This also appears to be in agreement with Fig.~\ref{fig:3}f which shows the local geometry at Wind swinging across $45^{\circ}$ throughout the two preceding hours. These obliquities, and the ordering above, use the L1 timing normal, which is stable within the fleet (Appendix~\ref{sec:appendix_diagnostics}); with the normal $\approx45^{\circ}$ away implied by longer-baseline timing (Appendix~\ref{sec:appendix_speed}), the interval would be predominantly quasi-perpendicular at all three spacecraft, further showing that no single normal characterizes the connection. The prominent high-frequency power at IMAP, in the whistler range \citep{Hoppe1983,wilson2012observations}, further points to actively evolving compressive structures upstream, which we examine next. However, because this frequency range lies close to the Nyquist frequency, its interpretation should be treated with caution.

\subsection{Shocklets: steepened waves far upstream}\label{sec:shocklets}

\begin{figure*}[ht]
    \centering
{\includegraphics[width=1\textwidth]{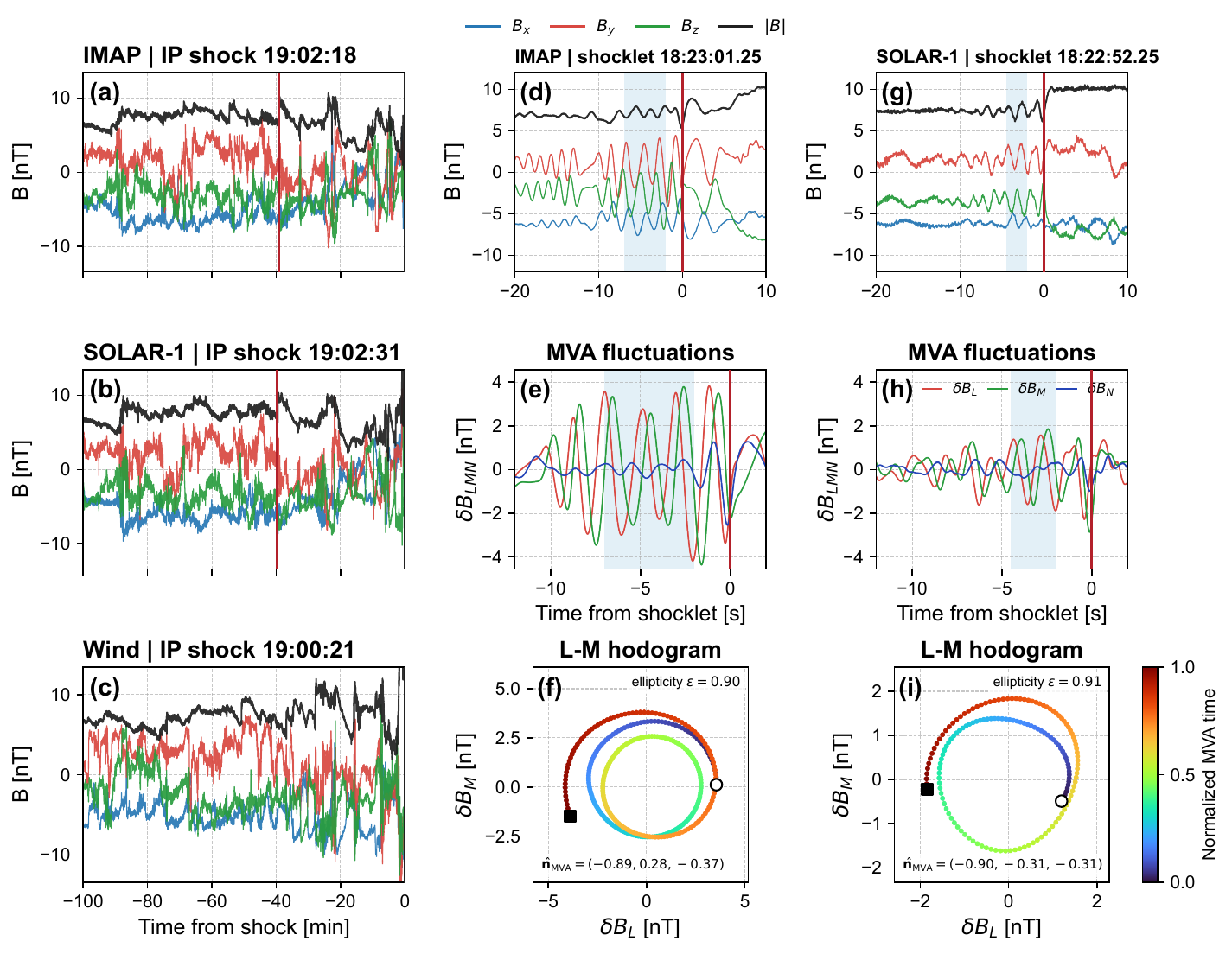}}\caption{L1 shocklet candidates and magnetic polarization. Panels (a)--(c) show 100 min of upstream GSE magnetic field at IMAP, SOLAR-1, and Wind relative to each local interplanetary-shock crossing. Red vertical lines identify approximate IMAP and SOLAR-1 shocklet ramps. Panels (d)--(f) and (g)--(i) show, respectively, the IMAP and SOLAR-1 candidates: the unfiltered native magnetic field, the $0.2$--$2.0$ Hz bandpass filtered magnetic field in MVA coordinates, and the corresponding $L$--$M$ hodogram. Pale shading marks the MVA intervals, $t-t_{\rm ramp}=[-7,-2]$ s (IMAP) and $[-4.5,-2]$ s (SOLAR-1), each containing a single wave packet. Hodogram color advances through the interval; the open circle and filled square mark its start and end. The ordered MVA eigenvalues $(\lambda_L,\lambda_M,\lambda_N)$ are $(5.114,4.120,0.050)\,\mathrm{nT^2}$ for IMAP and $(1.286,1.069,0.046)\,\mathrm{nT^2}$ for SOLAR-1. The shocklets exhibit compressive ramps and organized, nearly circular upstream magnetic rotations reflecting the presence of whistler precursors.} \label{fig:4}
\end{figure*}

Well upstream of the shock, the foreshock already contains discrete, steepened compressive structures. About forty minutes ahead of the ramp, nominally $\sim315\,R_E$ upstream, both IMAP and SOLAR-1 record an isolated compressive structure with an attached, oscillatory wave train (Fig.~\ref{fig:4}a,b; red markers), the typical signature of shocklets whose steepened ramps act locally as small quasi-perpendicular shocks \citep{Hoppe1981,wilson2009low,trotta2023multi,wilson2025large}. Minimum variance analysis of the band-passed field (Fig.~\ref{fig:4}d--i) shows the wave trains to be coherent and nearly circularly polarized, with ellipticities of $0.90$ at IMAP and $0.91$ at SOLAR-1, indicating compressive ramps with $B_d/B_u\approx1.5$. These are the organized, whistler precursors expected from steepening waves \citep{Scholer2003,Treumann2009,wilson2009low,wilson2017revisiting,trotta2023multi}. Wind, shown for context, observes similar structures but, given its larger distance from the other two spacecraft, is not assigned a directly comparable event. We note that the $\sim7\,R_E$ separation between IMAP and SOLAR-1 is small enough that both spacecraft likely sample the same individual shocklet rather than merely the same statistical population of steepened waves. This contrasts with previous multi-spacecraft shocklet observations at larger separations \citep{trotta2023multi}, where different structures were recorded at each monitor, and constitutes the first candidate two-point observation of the same shocklet.

Two features make these structures notable. First, they appear far outside the immediate shock transition and are seen at both spacecraft at nearly the same distance from their respective shocks, indicating that the steepening is a widespread property of the upstream medium rather than a local, single-spacecraft feature. While IMAP and SOLAR-1 are separated across the flow, in $Y$ and $Z$, and the exact orientation of the wave packet is unknown, their near-simultaneous appearance points to a common upstream wave population extended across the flow. Second, their location so far ahead of the shock is consistent with the steepening of pre-existing, ambient solar wind waves \citep{wilson2009low,trotta2023multi} rather than waves generated locally at the shock front \citep{Lucek1997,Wilson2007,Raptis2026a}.

\subsection{Candidate localized large-scale transients}\label{sec:transients}

\begin{figure*}[ht]
    \centering
{\includegraphics[width=1\textwidth]{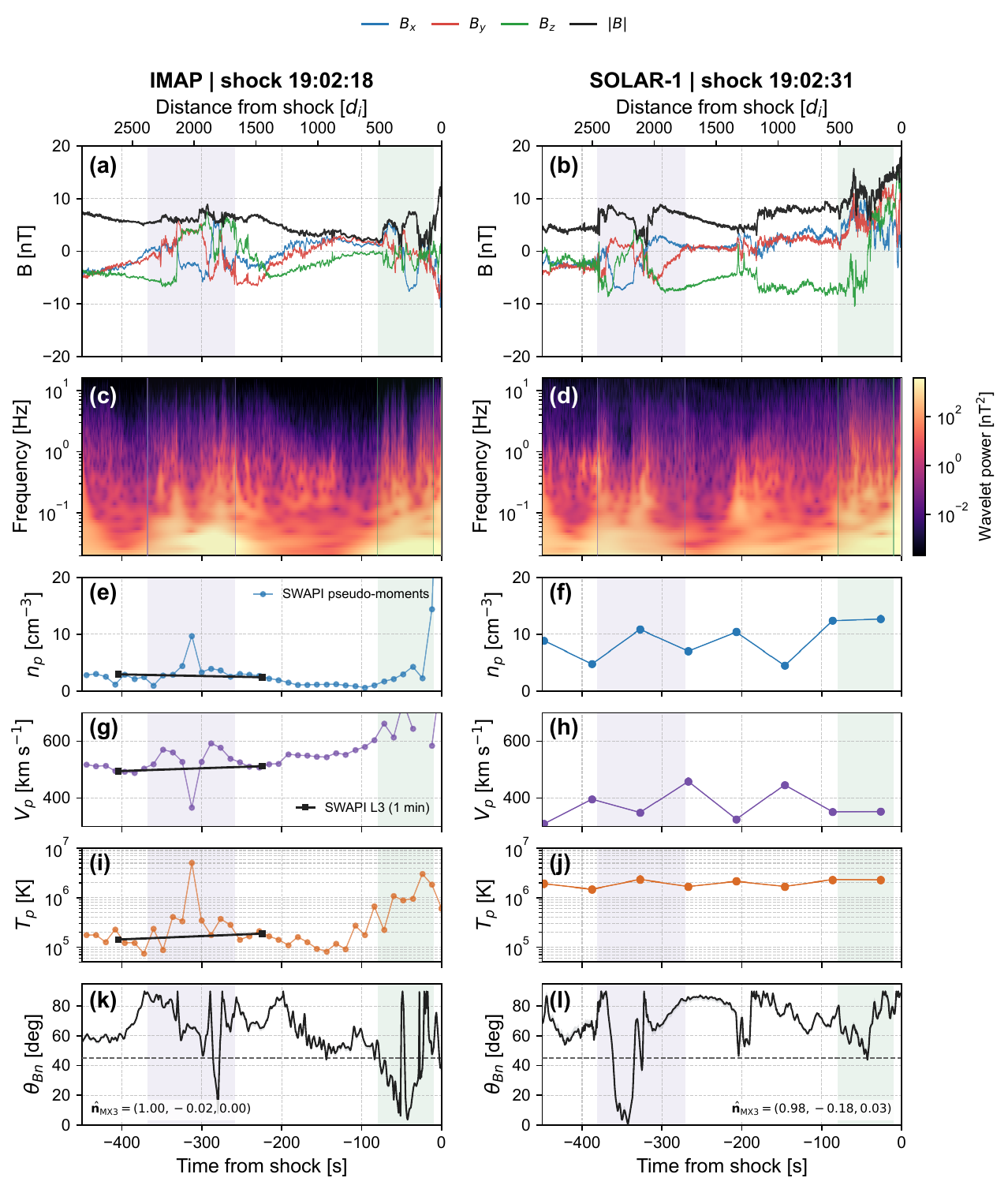}}\caption{{Upstream transient observations at IMAP and SOLAR-1. Left and right columns show IMAP and SOLAR-1 from 450~s upstream to each local interplanetary-shock crossing. Panels (a,b) show the native GSE magnetic components and magnitude; (c,d) the wavelet trace power spectra. Panels (e)--(j) show proton density, scalar speed, and temperature: colored points are unsmoothed IMAP/SWAPI pseudo-moments, black lines the 1-minute IMAP/SWAPI L3 moments, and SOLAR-1 shows the SWiPS L2 (uncorrected) moments. Panels (k,l) show $\theta_{Bn}$ obtained from a 3-s moving averaged field and the local MX3 normal (radial scalar-speed proxy), with gray shading spanning the 16th--84th percentile range and the dashed line at $45^{\circ}$. The two shaded bands (earlier, purple; later, green) mark localized interplanetary magnetic field (IMF) rotations, at roughly $18{:}56{:}10$--$18{:}58{:}00$~UTC and from 80 to 10~s before each local shock, respectively, that differ measurably between IMAP and SOLAR-1 and are candidate large-scale foreshock transients (e.g., HFAs or foreshock bubbles). Top axes give distance from the shock using $V_{\rm sh}=850\,\mathrm{km\,s^{-1}}$ and $d_i=131.6\,\mathrm{km}$.}}\label{fig:5}
\end{figure*}

Closest to the shock, within the final $\sim450$~s of upstream sampling, IMAP and SOLAR-1 each encounter localized structures with the character of large-scale foreshock transients (Fig.~\ref{fig:5}). Two intervals stand out at both spacecraft (shaded): each contains a localized rotation of the interplanetary field, accompanied by perturbed proton moments, a depressed and variable density
with localized heating, and a drop in the local shock obliquity toward quasi-parallel values (Fig.~\ref{fig:5}k,l). These signatures resemble those used to identify hot flow anomalies and foreshock bubbles at planetary bow shocks \citep{turner2013first,schwartz2000conditions,zhang_dayside_2022,Kajdi2024,Raptis2025b,Turc2025} and motivate our interpretation of the two intervals as candidate large-scale foreshock transients at an IP shock.

The notable point is that these structures differ between the two spacecraft. Although IMAP and SOLAR-1 are separated by only $\sim 7\,R_E$ ($\sim 340\,d_i$), the field rotations, plasma signatures, and obliquity depressions in each interval differ in amplitude and in internal structure. That separation is mostly
transverse to the flow, $(1.1,\,-6.4,\,-2.7)\,R_E$ in GSE, so a radially convecting structure reaches both spacecraft within seconds. The differences are therefore either spatial or, if temporal, very rapid. Spatially, these structures vary internally on scales comparable to the IMAP--SOLAR-1 separation ($\sim7\,R_E$), rather than appearing as
uniform features convected past two monitors. We cannot separate the two readings with this geometry, though transients at Earth have been shown to evolve on comparable timescales \citep{Liu2021,turner2021direct}. The two-point differences therefore show that the intervals are spatially structured and/or rapidly evolving on scales resolved by the IMAP--SOLAR-1 separation, rather than planar fluctuations convected past both spacecraft.

The morphology of the first candidate (purple shading) differs between the two spacecraft in a specific and suggestive way. At SOLAR-1 the structure shows more sharply steepened edges bounding a distinct core, with the field rotation concentrated in its interior, while at IMAP the same interval appears as a broader, less organized perturbation. These two morphologies bear a qualitative resemblance to different stages of the HFA evolutionary sequence at Earth's bow shock, in which an initially broad perturbation develops compressive edges and an interior cavity \citep{zhang_dayside_2022,Kajdi2024}, though the $\sim7\,R_E$ separation cannot rule out spatial structure within a single transient as an alternative explanation. The candidates span roughly 1--2~min (110 and 70~s), corresponding, at the ambient solar wind speed of $\sim500\,\mathrm{km/s}$, to spatial scales of $\sim270$--$420\,d_i$ ($\sim5$--$9\,R_E$), comparable to the candidates of bursty localized acceleration recently reported near shocks with Parker Solar Probe \citep{berland2025parker}. Whether this reflects their typical developed size or a snapshot of ongoing growth remains unclear. We label them as ``first candidates'' deliberately, as a firm classification will require analysis of more IP shocks and a search for associated downstream signatures at the near-Earth monitors, which we leave for future work. The second candidate is even more difficult to analyze because, while it shows very typical characteristics at IMAP, it appears to be already part of the shock transition layer by the time SOLAR-1 observes it. This contrast may indicate that the advancing shock overtook and absorbed the transient over the $\sim7\,R_E$ between the two spacecraft, implying rapid restructuring of the local shock transition on scales smaller than the transient itself.

These local departures notwithstanding, IMAP and SOLAR-1 agree well across the broader upstream interval, where both measure similar magnetic field and plasma properties (Fig.~\ref{fig:extended_upstream_context} and Appendix~\ref{sec:appendix_broader}). The exceptions are the proton density and
temperature near the shock (Fig.~\ref{fig:5}e,f,i,j), which differ significantly between the two spacecraft. Because both instruments were early in commissioning during this event, part of this discrepancy likely reflects instrumental effects at both spacecraft, including possible contamination of the SWAPI and SWiPS proton moments by penetrating energetic ions in the foreshock environment. Together with the 1-minute cadence of the SOLAR-1 moments and the 12-s IMAP pseudo-moments, these offsets prevent us from confirming whether the density and temperature perturbations fully match the expected transient signatures. We therefore base the transient identification primarily on the magnetic field variability, which is measured independently and at high cadence by both spacecraft.

\section{IP Shock Evolution: Sheath Growth from L1 to MMS}\label{sec:sheath}

\begin{figure*}[ht]
    \centering
{\includegraphics[width=1\textwidth]{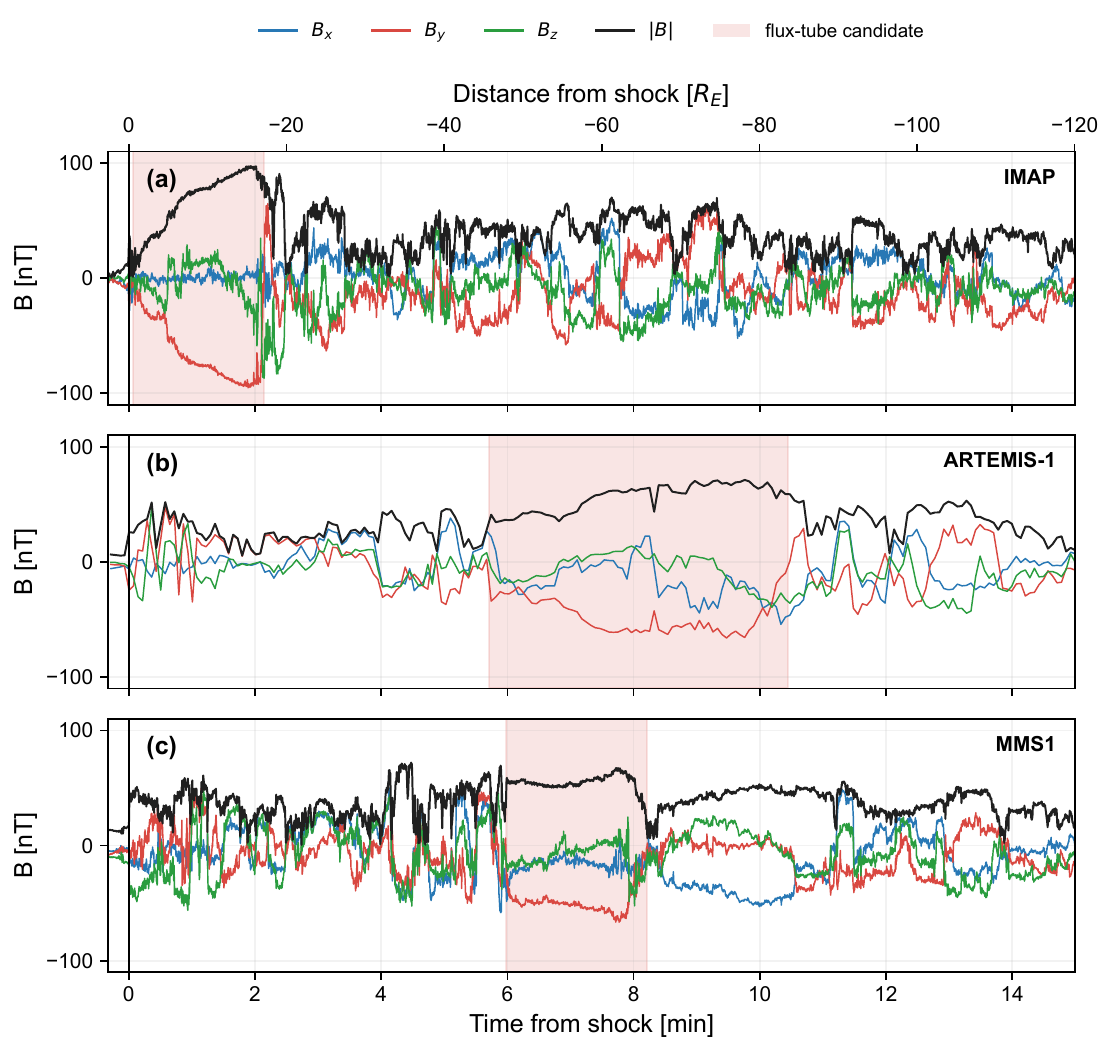}}\caption{Evolution of the downstream flux-tube envelope from IMAP to MMS1. Panels (a)--(c) show the native GSE magnetic field components and magnitude at IMAP, ARTEMIS-1, and MMS1 from 20 s before to 15 min after each local interplanetary-shock crossing. Black vertical lines mark the local shocks. Red shading marks the broad flux-tube envelope, bounded by the $B_y$ sign changes around its negative $B_y$ interval, highlighting its relative location with respect to the shock. The upper axis gives effective downstream distance using $V_{\rm sh}=850\,\mathrm{km/s}$ and $R_E=6371\,\mathrm{km}$. Based on this, the shock-to-envelope separation grows from $<1\,R_E$ at IMAP to $\approx46\,R_E$ at ARTEMIS-1 and $\approx48\,R_E$ at MMS, just upstream of Earth's bow shock.} \label{fig:6}
\end{figure*}

The same flux tube that influences the shock parameters at L1 also provides a timing marker for tracking the downstream evolution toward Earth. Between the shock ramp and the leading edge of the flux tube lies the sheath plasma shocked since the flux tube was overtaken, so its growth is simply the change in the relative downstream distance between the two boundaries. Figure~\ref{fig:6} compares this separation at IMAP, ARTEMIS-1, and MMS. At IMAP the flux tube sits immediately behind the shock (Fig.~\ref{fig:6}a), whereas at ARTEMIS-1 and MMS the same structure, identified by a matching coherent $B_y$ rotation (Fig.~\ref{fig:6}b,c), trails the shock by roughly six minutes, or $\sim50\,R_E$. Over the $\sim200\,R_E$ from L1 to near-Earth space, the shock-to-flux-tube separation has therefore grown by about $50\,R_E$ along the Sun-Earth line. ARTEMIS-1, about 75\% of the way from IMAP to MMS, already records a lag of $\approx343$~s ($\approx46\,R_E$), compared with $\approx358$~s ($\approx48\,R_E$) at MMS, so most of this growth occurred before lunar orbit, consistent with the faster shock transit between L1 and lunar orbit (Appendix~\ref{sec:appendix_speed}); its broader flux-tube profile suggests a different path through the structure.

This growth is quantitatively consistent with the transit times of the two boundaries. Between IMAP and MMS, the shock advances at a mean $\approx1390\,\mathrm{km/s}$ along $\hat{\mathbf n}_{\rm timing}$ (Section~\ref{sec:overview}), whereas the leading edge of the flux tube travels at $\approx975\,\mathrm{km/s}$ between IMAP and ARTEMIS-1 and $\approx1010\,\mathrm{km/s}$ between ARTEMIS-1 and MMS, consistent with the downstream flow (Figure~\ref{fig:speed}). The shock thus outruns the flux tube at $\approx410\,\mathrm{km/s}$, and over the $\sim850$~s from IMAP to MMS this widens the sheath by $\approx55\,R_E$, consistent with the $\approx50\,R_E$ separation at MMS. These two consistent estimates, the direct envelope timing and the difference between the shock and flux-tube-edge speeds, agree with each other, making the result more robust.

The flux tube itself changes between L1 and Earth (median $B_y\approx-75$, $-54$, and $-50$~nT at IMAP, ARTEMIS-1, and MMS), as does the field behind it (Figure~\ref{fig:6}). Besides different crossing paths, such changes can result from the evolution of the downstream turbulence, which proceeds on the nonlinear (eddy-turnover) time $\tau_{\rm nl}\sim\lambda/\delta v$, i.e., the time for fluctuations of scale $\lambda$ and velocity amplitude $\delta v$ to transfer their energy to smaller scales. Taking $\lambda$ from the magnetic correlation time and the downstream flow speed, and $\delta v$ from the magnetic fluctuation amplitude, we obtain $\tau_{\rm nl}\sim1$--$2$~min (at most $\sim6$~min for the longest windows). This is well below the $\approx20$~min that elapse between the IMAP and MMS observations of the same plasma, so the downstream fluctuations have had several nonlinear times to evolve before reaching Earth, although spatial and temporal variations cannot be fully separated with these observations.

The sheath demonstrably grows between L1 and Earth, so an upstream monitor samples a thinner sheath than the one that eventually reaches the magnetosphere \citep{Lugaz2024}. Although this thickening corresponds to only about six minutes of sheath crossing at Earth, it represents a substantial increase in the volume of shocked, geoeffective plasma that the magnetosphere encounters.

\begin{figure*}[ht]
    \centering
    {\includegraphics[width=1\textwidth]{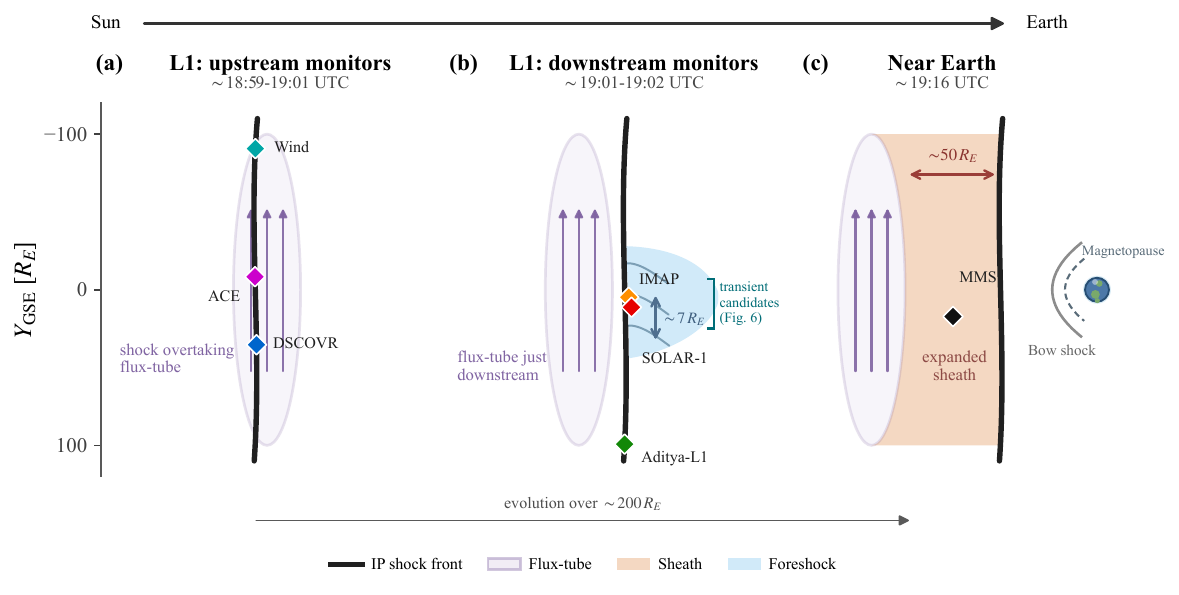}}
\caption{Schematic of the 2026 January 19 interplanetary shock, viewed in the $X$--$Y$ GSE (ecliptic) plane. The flux-tube cross-section is illustrated with arrows denoting its strong $-B_y$ component. (a)~At the upstream L1 monitors ($\sim$18:59--19:01~UTC), the shock front is overtaking the flux tube; ACE, Wind, and DSCOVR sample it both upstream and downstream. (b)~At the downstream L1 monitors ($\sim$19:01--19:02~UTC), the shock has just passed through the flux tube, which now lies immediately downstream. IMAP and SOLAR-1, separated by $\sim7\,R_E$ ($\sim340\,d_i$), sample an active upstream foreshock that hosts the candidate transient signatures analyzed in Figure~\ref{fig:5}. The inflated compression ratios at these spacecraft (Figure~\ref{fig:diagnostics}) result from the flux tube entering the downstream averaging interval. (c)~Near Earth ($\sim$ 19:16~UTC), the shock has outrun the flux tube by approximately $50\,R_E$, and MMS observes the expanded sheath. Spacecraft $Y_{\rm GSE}$ positions follow Figure~\ref{fig:1}; the orientation of the flux tube relative to the shock front is symbolic; the data constrain its extent but not its orientation with respect to the shock surface. The flux tube is drawn as a uniform envelope for simplicity. All other spatial relationships are illustrative and not to scale.} \label{fig:synthesis}
\end{figure*}

\section{Discussion}\label{sec:discussion}

The measurements presented in Sections~\ref{sec:mhd}--\ref{sec:sheath} follow the shock, its foreshock, and its sheath as one coupled system across kinetic, mesoscale, MHD, and global system scales (Figure~\ref{fig:synthesis}). Here we discuss what each result implies for shock physics and space-weather modeling, and identify the open questions that multipoint L1 observations can now address.

First, the flux tube biasing shows how difficult the near-shock geometry can be to determine from one or a few spacecraft. Its presence inflated the apparent compression ratio beyond the strong-shock limit and split the single-spacecraft and multi-spacecraft obliquity estimates at IMAP and SOLAR-1. A system-size evaluation across the six L1 monitors resolved the ambiguity and identified a distinct flux tube as the cause, yet it also exposed how limited standard shock-geometry techniques remain once the upstream and downstream intervals can no longer be assumed to sample the same plasma. A promising direction is to develop this multipoint approach further and to treat the variability of the upstream and downstream wave field and particle distributions as a more physical determinant of the shock geometry (see Figure~\ref{fig:3}), rather than relying on jump conditions that assume a laterally uniform, quasi-stationary shock. It should be noted that while the exact nature of this solar wind flux tube is outside the scope of this article, its properties and scale sizes resemble flux ropes routinely observed in the solar wind \citep{borovsky2008flux,ruan2009multiple,blanco2019multispacecraft,shin2026statistical}. More broadly, the interaction of flux tubes with shocks can have direct implications for particle acceleration. When individual flux tubes meet the shock surface, segments of the same field line can lie simultaneously upstream and downstream. Particles propagating along such field lines undergo repeated shock crossings and gain energy through a process related to DSA, even at nominally perpendicular geometries \citep{Zank2006}. Indeed, \citet{giacalone2021energetic} showed that when an IP shock crosses a large pre-existing magnetic structure in the solar wind, the interaction can produce local variations in which energetic particles are accelerated.

Second, the active foreshock resolved at IMAP and SOLAR-1 has significant reach. Hundreds of $R_E$ ahead of the shock, pre-existing solar wind fluctuations were already steepening into shocklets, supporting recent reports at other IP shocks \citep{trotta2023multi}. This shows that localized wave steepening can occur far upstream within IP-shock foreshocks and, through the attendant high-frequency whistler activity, establish the wave environment in which suprathermal particles are scattered and pre-accelerated. Such structures appearing far ahead of the shock broadens the region over which an IP shock can influence, and be influenced by, its upstream medium.

Third, near the shock, we reported the first candidate signatures of large-scale foreshock transients at an IP shock, with field, plasma, and obliquity signatures resembling the foreshock bubbles and hot flow anomalies routinely seen at planetary bow shocks. Their scales, of roughly 5--10~$R_E$ (several hundred $d_i$), exceed those of their terrestrial counterparts but not by a large factor. This is broadly consistent with transient size expanding with the host shock \citep{zhang_dayside_2022,Kajdi2024,Raptis2026b}, though the modest size may instead reflect insufficient time to develop by the moment of observation, or a fundamental limit on their growth set by the motion of the shock relative to the IMF rotation that seeds them. The localized IMF rotations observed within both candidate intervals (Figure~\ref{fig:5}) are the most direct candidates for the seeding discontinuities; the tangential discontinuity bounding the flux tube, which meets the shock near IMAP and SOLAR-1 at about the time of the crossing, could additionally contribute to the second candidate. These are conjectures: distinguishing them, like establishing the transients themselves, will require dedicated modeling and further events, alongside study of the particle energization they drive. At Earth, such transients can retain part of their structure across the shock, drive clear energization in the magnetosheath, and cause subsequent downstream transient structures \citep{lin2002global, vsafrankova2002structure, Raptis2025b,Zhou2026,guo2026large}. Whether the same holds here, in particular in terms of particle acceleration, is the natural focus of a follow-up study.

Finally, the shock's evolution from L1 to Earth is itself significant. While CMEs and their fronts (IP shocks) are known to evolve throughout the heliosphere \citep{Kilpua2017,manchester2017physical,luspay2026structure}, the evolution from L1 to Earth remains much less systematically explored. Roughly $50\,R_E$ of shocked plasma accumulated downstream of the shock between IMAP and MMS, extending the sheath appreciably (Figs.~\ref{fig:6} and \ref{fig:synthesis}c). The flux tube and the fluctuations behind it also changed over this distance, consistent with the evolution and delayed decay of turbulence behind IP shocks \citep{Pitna2017,Pitna2023,Pitna2024}. Although this expansion unfolds on temporal scales much shorter than the development of a storm main phase, it shows that an upstream monitor and the magnetosphere can encounter measurably different sheaths. This argues for incorporating near-Earth observations, such as from MMS, THEMIS, or ARTEMIS, into data-driven modeling of the L1-to-Earth interval during such events. It also suggests that machine-learning models of near-Earth solar wind conditions \citep{OBrien2023} might improve by taking multiple L1 monitors as inputs and multiple near-Earth monitors as targets. The same reasoning extends to associating upstream solar drivers with their space-weather effects, including the variability of the radiation belts \citep{pandya2019variation,capannolo2026solar}, the excitation of magnetospheric waves \citep{Liu2025foreshock,Wing2026}, and solar wind-magnetosphere coupling schemes that incorporate upstream conditions \citep{Smirnov2020,Hu2022,hua2026neural}. Finally, the same multipoint framing could improve the driving of global magnetosphere simulations \citep{toth2005space,zhang2019gamera}, so that what drives the simulation better resembles what actually reaches Earth's magnetosphere.

Several research objectives follow directly from this event. The most immediate is to characterize the flux tube in greater detail as it is overtaken by the shock, quantifying how it steepens, how it couples to the shock, and whether the interaction with reflected and energetic particles modifies the flux-tube structure itself. Recent work on the transmission of small-scale magnetic islands and flux ropes through perpendicular shocks \citep{zank2021flux,Nakanotani2022} can provide the framework needed for such an investigation.

In parallel, tracking the candidate transients and shocklets into the downstream region using near-Earth monitors, and searching for their signatures and any associated particle energization, would test how much of the upstream transient activity survives to reach geospace. At global system scales, the space-weather implications of a highly variable CME sheath could be assessed through modeling by comparing simulations driven by near-Earth observations with those driven by L1 observations. Such a comparison would help quantify the extent to which sheath variability and its evolution between L1 and Earth can influence the accuracy of geospace modeling.

\section{Conclusion}\label{sec:conclusion}

By analyzing the multi-scale evolution of a fast IP shock on 2026 January 19, we obtain the following main findings:

\begin{enumerate}
    \item A mesoscale solar wind flux tube, overtaken by the shock during
    its passage across the L1 fleet, biased single-spacecraft compression ratios up to six times the strong-shock limit and generated an apparent variability in the normal direction. The multi-spacecraft comparison resolved the ambiguity that coplanarity-based methods
    alone could not.

    \item IMAP and SOLAR-1 recorded the first candidate signatures of large-scale foreshock transients at an interplanetary shock. The structures differ measurably between the two spacecraft despite their $\sim7\,R_E$ separation, indicating spatial inhomogeneity
    or temporal variability on scales comparable to the structures themselves.

    \item Steepened shocklets appear nominally $\sim315\,R_E$ upstream of the shock at both IMAP and SOLAR-1, consistent with the steepening of pre-existing solar wind waves rather than local generation at the
    shock front.

    \item The shock-to-flux-tube separation grew by approximately $50\,R_E$ between L1 and ARTEMIS-1/MMS, tracked using the flux tube as a downstream boundary marker and indicating nominal sheath thickening by the same amount. An upstream monitor therefore samples a thinner,
    different sheath than the one that reaches the magnetosphere.
\end{enumerate}

Together, these observations show that an interplanetary shock, its foreshock, and the sheath behind it evolve measurably across scales and along the Sun-Earth line. The emerging multipoint L1 fleet, alongside the near-Earth observatories, now makes it possible to resolve this evolution and study IP shocks and their interaction with Earth's geospace as a coupled cross-scale system. 

\section*{Acknowledgments}

\begin{acknowledgments}

We thank NASA’s IMAP mission team for their contributions to the design, development, commissioning, and validation of the IMAP science and data products. This research effort was supported and funded by the IMAP mission (80GSFC19C0027), as part of NASA’s STP mission line. IMAP magnetometer operations are funded by the UK Space Agency (grant UKRI2494). SR additionally acknowledges the NSF GEM Focus Group ``Multiscale Dayside Transients and Their Effect on Earth's Magnetosphere''. The work of DC is supported by the Department of Space, Government of India. TH is supported by STFC grant UKRI1202. We thank the SuperMAG collaborators and the THEMIS/ARTEMIS team. \\

IMAP data are publicly available at the Space Physics Data Facility
\url{https://cdaweb.gsfc.nasa.gov/} or through \url{https://imap.princeton.edu/data}. MMS data can be found through the public SDC \url{https://lasp.colorado.edu/mms/sdc/public/about/browse-wrapper/} or the Graphical User Interface (GUI) found in \url{https://lasp.colorado.edu/mms/sdc/public/search/}. The location, magnetic field, and plasma measurements of the Wind, ACE, ARTEMIS, and DSCOVR missions were taken from Coordinated Data Analysis Web (CDAWeb) \url{https://cdaweb.gsfc.nasa.gov}. The SMR index is available from SuperMAG at \url{https://supermag.jhuapl.edu/}. Aditya-L1 data are openly accessible via ISRO Science Data Archive (ISDA) \url{https://pradan1.issdc.gov.in/al1/}. The SOLAR-1 magnetic field measurements used in this study are pre-provisional MAG data made available by the SWFO/SOLAR-1 team. SOLAR-1 operational MAG data are publicly accessible through the NOAA Space Weather Prediction Center (SWPC; \url{https://www.spaceweather.gov/products/solar-wind}). Upon completion of validation, the finalized SOLAR-1 datasets are available through the NOAA/National Centers for Environmental Information (NCEI) Space Weather Portal (SPOT; \url{https://www.ncei.noaa.gov/products/space-weather}).
\end{acknowledgments}

\vspace{5mm}

\software{Python 3.12, NumPy, SciPy, pandas, Matplotlib, cdflib, and PySPEDAS. PSDs were computed using \texttt{scipy.signal.welch} (v1.13).}

\appendix

\section{Extended upstream context at IMAP and SOLAR-1}\label{sec:appendix_broader}

\begin{figure*}[ht]
    \centering
{\includegraphics[width=1\textwidth]{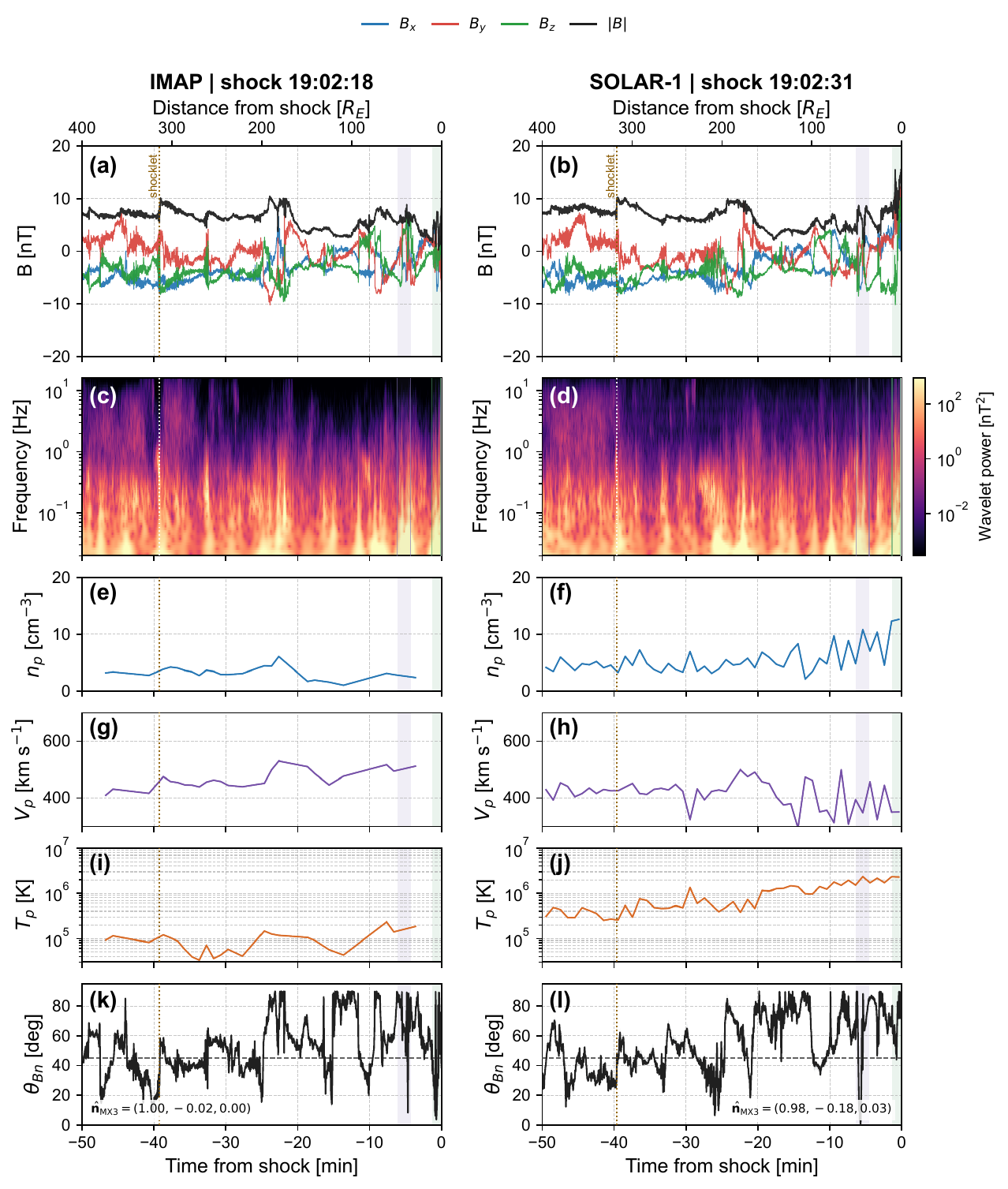}}\caption{Extended upstream context at IMAP and SOLAR-1. Same format and conventions as Figure~\ref{fig:5}, extended from 50 min upstream to each local shock crossing. Magnetic field and $\theta_{Bn}$ calculations use the native 64-Hz MAG cadence and the plotted traces retain every 16th native sample; the displayed Morlet trace power is averaged into non-overlapping 0.25-s bins after native-cadence wavelet calculation. Proton moments are shown as lines at their native sampling and contain only measured samples strictly upstream of each shock; proton temperature is logarithmic. Pale purple and green regions mark two localized IMF rotations, respectively, that appear to differ measurably between IMAP and SOLAR-1 and are potential large-scale foreshock-transient candidates (e.g., HFAs or foreshock bubbles). Thin gold intervals and dotted lines mark the MVA wave packets and approximate ramps of the IMAP and SOLAR-1 shocklet candidates analyzed in Figure~\ref{fig:4}. The upper axes show effective distance in Earth radii using $V_{\rm sh}=850\,\mathrm{km/s}$ and $R_E=6371\,\mathrm{km}$.}\label{fig:extended_upstream_context}
\end{figure*}

Figure~\ref{fig:extended_upstream_context} places the transient and shocklet candidates discussed in the main text within 50~min upstream of the local IMAP and SOLAR-1 shock crossings. Over this broader interval, the two spacecraft measure closely comparable magnetic field magnitudes and components and comparable proton bulk speeds, confirming that they sample the same large-scale solar wind stream and foreshock environment. Proton density and temperature show systematic offsets between the two instruments, discussed in Section~\ref{sec:transients}, and we therefore do not use them for cross-spacecraft comparison. The wavelet-power and MX3-based $\theta_{Bn}$ panels provide the local field--shock geometry and support that both spacecraft share the same shock geometry for most of the interval.

\section{Shock Timing and Diagnostic Values from the L1 Constellation}\label{sec:appendix_diagnostics}

Table~\ref{tab:l1_mc_appendix} lists the full numerical shock diagnostics summarized visually in Figure~\ref{fig:diagnostics}. The coplanarity normal vectors, angular deviations, and compression ratios with their 16th--84th percentile sensitivity ranges are shown here for quantitative comparison with other events. The median upstream field magnitudes at IMAP (5.6~nT) and SOLAR-1 (5.9~nT) are comparable to ACE (5.7~nT) and DSCOVR (6.3~nT), confirming that the inflated compression ratios at IMAP and SOLAR-1 arise from the unique downstream interval (see Section~\ref{sec:mhd}) rather than from anomalously low upstream fields. The broader spread across the fleet (5.6--9.1~nT) itself reflects the spatial mesoscale variability in the upstream solar wind.

\begin{table*}
\centering
\caption{Shock diagnostics at the six L1 spacecraft. The timing obliquities use a common six-spacecraft timing normal $\hat{\mathbf n}_{\rm timing}=(-0.964,-0.127,-0.235)$ in GSE coordinates. Bracketed values give the 16th--84th percentile ranges across the upstream/downstream window ensemble. The tabulated $B_u$ values are broader-context upstream medians computed over $-30$ to $-10$~min and are distinct from the upstream field averages used to calculate the compression ratios. See also Figure~\ref{fig:diagnostics}}
\label{tab:l1_mc_appendix}
\small
\setlength{\tabcolsep}{3.8pt}
\renewcommand{\arraystretch}{1.25}
\begin{tabular}{lcccccccc}
\hline\hline
S/C & \shortstack{Shock\\(UTC)} & \shortstack{$\hat{\mathbf n}_{\rm MC}$ (GSE)\\$(n_x,n_y,n_z)$} & \shortstack{$\Delta n$\\(deg)} & \shortstack{$\theta_{Bn}^{\rm MC}$\\(deg)} & \shortstack{$\theta_{Bn}^{\rm timing}$\\(deg)} & \shortstack{$B_u$\\(nT)} & $B_d/B_u$ & $B_{\rm peak}/B_u$ \\
\hline
ACE & 18:58:45 & $(-0.998,-0.036,-0.054)$ & \shortstack{$38.7$\\{\footnotesize$[18.2,62.0]$}} & \shortstack{$81.0$\\{\footnotesize$[69.5,86.1]$}} & \shortstack{$69.8$\\{\footnotesize$[66.4,71.0]$}} & $5.7$ & \shortstack{$4.54$\\{\footnotesize$[4.37,4.70]$}} & \shortstack{$5.15$\\{\footnotesize$[5.01,5.26]$}} \\
Wind & 19:00:21 & $(-0.709,0.049,-0.703)$ & \shortstack{$5.6$\\{\footnotesize$[3.1,8.0]$}} & \shortstack{$43.7$\\{\footnotesize$[38.8,50.0]$}} & \shortstack{$42.2$\\{\footnotesize$[39.9,44.0]$}} & $8.3$ & \shortstack{$3.19$\\{\footnotesize$[2.77,3.60]$}} & \shortstack{$4.41$\\{\footnotesize$[4.12,5.00]$}} \\
DSCOVR & 19:01:00 & $(-0.990,-0.097,-0.101)$ & \shortstack{$12.5$\\{\footnotesize$[4.4,24.4]$}} & \shortstack{$66.9$\\{\footnotesize$[62.5,71.2]$}} & \shortstack{$63.9$\\{\footnotesize$[60.6,65.4]$}} & $6.3$ & \shortstack{$3.49$\\{\footnotesize$[3.39,3.56]$}} & \shortstack{$3.70$\\{\footnotesize$[3.65,3.73]$}} \\
Aditya-L1 & 19:01:12 & $(-0.628,-0.257,0.734)$ & \shortstack{$6.0$\\{\footnotesize$[3.3,12.5]$}} & \shortstack{$22.8$\\{\footnotesize$[6.2,37.1]$}} & \shortstack{$71.2$\\{\footnotesize$[62.2,80.3]$}} & $9.1$ & \shortstack{$3.23$\\{\footnotesize$[2.68,4.33]$}} & \shortstack{$6.72$\\{\footnotesize$[6.46,7.11]$}} \\
IMAP & 19:02:18 & $(-0.747,-0.049,0.664)$ & \shortstack{$12.4$\\{\footnotesize$[6.8,35.5]$}} & \shortstack{$36.2$\\{\footnotesize$[25.4,44.6]$}} & \shortstack{$63.4$\\{\footnotesize$[48.6,76.5]$}} & $5.6$ & \shortstack{$23.39$\\{\footnotesize$[16.11,26.54]$}} & \shortstack{$29.22$\\{\footnotesize$[19.81,33.13]$}} \\
SOLAR-1 & 19:02:31 & $(-0.631,0.068,0.773)$ & \shortstack{$9.3$\\{\footnotesize$[5.6,13.9]$}} & \shortstack{$34.5$\\{\footnotesize$[24.7,45.9]$}} & \shortstack{$61.6$\\{\footnotesize$[51.2,67.1]$}} & $5.9$ & \shortstack{$8.06$\\{\footnotesize$[7.07,8.49]$}} & \shortstack{$10.70$\\{\footnotesize$[9.39,11.19]$}} \\
\hline
\end{tabular}
\begin{flushleft}
\footnotesize Each two-line diagnostic cell gives the median on the first line and the 16th--84th percentile range below it. $B_u$ is the median upstream $|B|$ computed over $-30$ to $-10$~min relative to each local shock crossing. $\Delta n$ is the angular deviation of the magnetic coplanarity (MC) normals from their consensus direction. The timing column uses the common six-spacecraft timing normal and each spacecraft's local upstream field. The final column reports peak rather than median compression. Central values are medians over varied upstream and downstream intervals within $-180$ to $-20$~s and $+15$ to $+180$~s, respectively.
\end{flushleft}
\end{table*}

The common timing normal in Table~\ref{tab:l1_mc_appendix} is obtained from the ramp times in Table~\ref{tab:timing_audit}, identified at the steepest increase of $|B|$ after a 3-s median filter and manually verified. Minimizing $\sum_i e_i^2$ for $t_i=t_0+\mathbf r_i\cdot\mathbf s+e_i$ over $t_0$ yields Equation~\ref{eq:timing_objective}, solved by least squares with equal weights. The residual RMS is 5.0~s; referencing the fit to the IMAP crossing gives $893\,\mathrm{km/s}$ and a normal within $0.8^{\circ}$, and omitting one spacecraft at a time gives a range of 820--949~km/s. The main caveats are the assumption of a planar front at constant speed across the $\sim200\,R_E$ aperture of the fleet, which averages over any inclination or rippling, and ramp-identification errors, which map directly into the residuals; the largest residual occurs at SOLAR-1, where the second transient candidate merges with the shock (Section~\ref{sec:transients}).

\begin{table}[ht]
\centering
\caption{L1 ramp times, residuals $e_i$ of the full timing fit, and the shock speed obtained when the listed spacecraft is omitted.}
\label{tab:timing_audit}
\begin{tabular}{lccc}
\hline\hline
S/C & UTC & $e_i$ (s) & $V_{\rm sh}$ without S/C (km/s) \\
\hline
ACE & 18:58:45 & $+2.51$ & 819.9 \\
Wind & 19:00:21 & $-2.32$ & 865.1 \\
DSCOVR & 19:01:00 & $-1.99$ & 840.9 \\
Aditya-L1 & 19:01:12 & $-2.00$ & 948.7 \\
IMAP & 19:02:18 & $-5.94$ & 825.1 \\
SOLAR-1 & 19:02:31 & $+9.74$ & 885.9 \\
\hline
\end{tabular}
\end{table}

\section{Magnitude and Trace Spectra}\label{sec:appendix_spectra}

\begin{figure*}[ht]
\centering
\includegraphics[width=\textwidth]{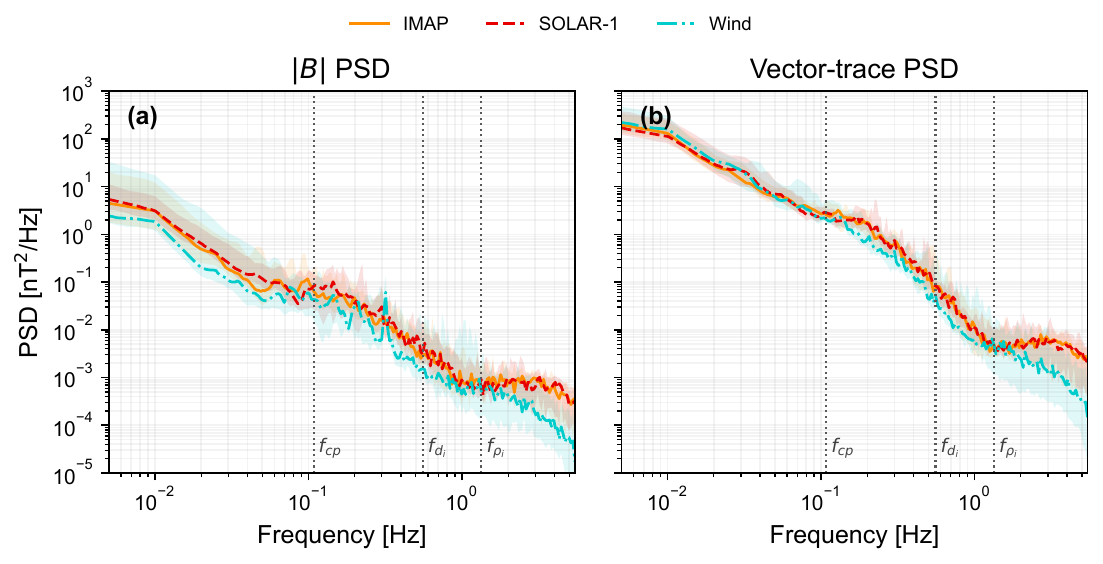}
\caption{Upstream (a) $|B|$ and (b) trace PSDs at IMAP, SOLAR-1, and Wind, using the same 30-min windows as Figure~\ref{fig:3}. Lines are frequency-wise medians over the windows and shading spans their 16th--84th percentiles, i.e., the temporal variability over the two hours preceding each shock. Dotted lines mark the proton cyclotron frequency $f_{cp}$ and the spacecraft-frame frequencies $f_{d_i}=V_{\rm sw}/(2\pi d_i)$ and $f_{\rho_i}=V_{\rm sw}/(2\pi\rho_i)$ of the proton inertial length and thermal gyroradius, from the median upstream IMAP and Wind moments.}
\label{fig:revision_spectra}
\end{figure*}

Figure~\ref{fig:revision_spectra} compares the median $|B|$ and trace PSDs over the windows of Figure~\ref{fig:3}. Over $0.01$--$0.1$~Hz, the median IMAP-to-Wind power ratio is 1.60 for $|B|$ but 0.86 for the trace, so the compressibility $P_{|B|}/P_{\rm trace}$ is $\approx0.02$ at IMAP and SOLAR-1 but $\approx0.01$ at Wind. Between 0.1 and 1~Hz, both ratios are $\approx1.6$ and the compressibility is similar at all three spacecraft; the excess persists over 1--3~Hz, below Wind's near-Nyquist range (1.32 and 1.67). SOLAR-1 has the same ULF $|B|$ power as IMAP, so this excess is not an IMAP instrumental effect. All spectra flatten between $\approx0.06$ and 0.15~Hz, close to the proton cyclotron frequency ($f_{cp}\approx0.1$~Hz) and to the spacecraft-frame frequency of ULF waves with wavelengths of $\sim1\,R_E$ ($V_{\rm sw}/R_E\approx0.07$~Hz), and steepen most strongly near the spacecraft-frame frequency of the proton inertial length ($f_{d_i}\approx0.6$~Hz). Above that of the thermal gyroradius ($f_{\rho_i}\approx1.3$~Hz), IMAP and SOLAR-1 show a second bump near 2--3~Hz, in the whistler range (Section~\ref{sec:kinetic}). The IMAP and SOLAR-1 excess between 0.1 and 1~Hz thus lies between the ULF and ion-kinetic scales, consistent with stronger foreshock wave activity, and the spectral shape resembles foreshock spectra at Earth's bow shock \citep{Eastwood2005,Raptis2026a,Zhao2026}; identifying the waves involved requires a wave--particle resonance analysis, which we leave for future work.

\section{Solar Wind Speed and Shock Propagation Speed}\label{sec:appendix_speed}

\begin{figure*}[ht]
\centering
\includegraphics[width=\textwidth]{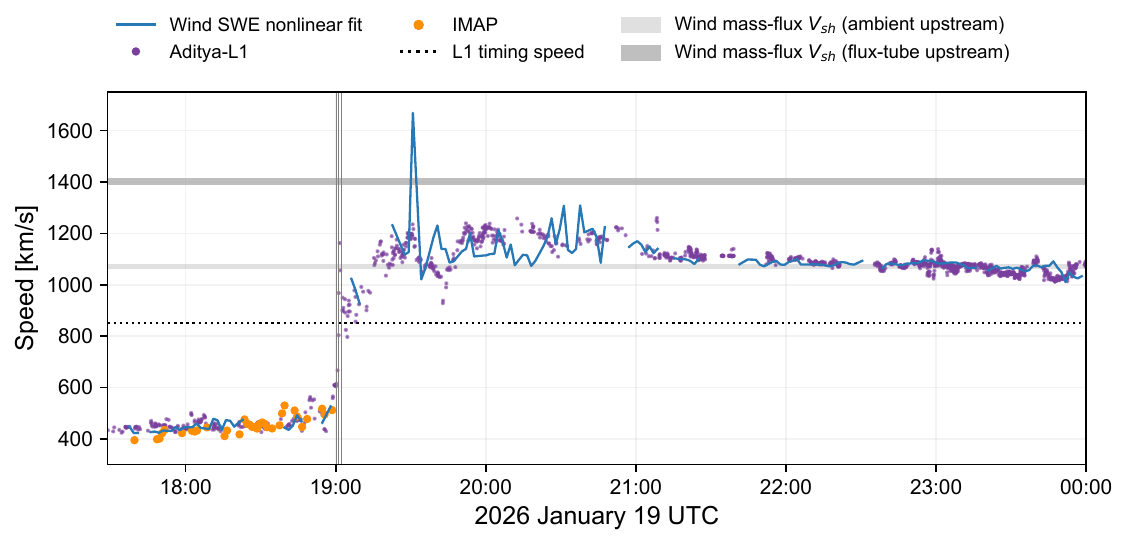}
\caption{Proton speed at L1 between 17:28:45 and 24:00:00 UTC from Wind SWE (nonlinear fits), Aditya-L1, and IMAP. Vertical lines mark the local shocks at Wind, Aditya-L1, and IMAP. The dotted line marks the six-spacecraft L1 timing speed ($852\,\mathrm{km/s}$). Gray bands show the local shock speeds at Wind from mass-flux conservation using the ambient solar wind ($1060$--$1080\,\mathrm{km/s}$, lower band) and the flux-tube plasma ($1390$--$1410\,\mathrm{km/s}$, upper band) as the upstream state.}
\label{fig:speed}
\end{figure*}

Figure~\ref{fig:speed} shows the available proton speeds from Wind SWE \citep{Ogilvie1995}, Aditya-L1 ASPEX/SWIS \citep{goyal2018aditya}, and IMAP (upstream only); ACE SWEPAM speeds are invalid during the event and DSCOVR plasma data were not available. The local shock speed at Wind from mass-flux conservation, $V_{\rm sh}=(N_dV_{d,n}-N_uV_{u,n})/(N_d-N_u)$ with $V_n=\mathbf V\cdot\hat{\mathbf n}_{\rm timing}$, depends on the upstream state: $\approx1060$--$1080\,\mathrm{km/s}$ with the ambient solar wind and $\approx1390$--$1410\,\mathrm{km/s}$ with the flux-tube plasma directly ahead of the ramp, although each window holds only a few SWE samples. The shock arrival times give mean speeds along $\hat{\mathbf n}_{\rm timing}$ of $\approx1290$--$1590\,\mathrm{km/s}$ from the L1 spacecraft to ARTEMIS-1 or MMS and $\approx1070\,\mathrm{km/s}$ from ARTEMIS-1 to MMS. No single plane reproduces all crossings (residuals up to $\approx40$~s), and a plane through Wind, IMAP, SOLAR-1, and ARTEMIS-1 requires $\approx1470\,\mathrm{km/s}$ with a normal $\approx45^{\circ}$ from $\hat{\mathbf n}_{\rm timing}$.

\end{document}